\documentclass[11pt, a4paper]{article}
\usepackage{import}
\usepackage{example}
\usepackage[english]{babel}
\usepackage{adjustbox}
\usepackage{array}
\usepackage{amsthm}
\usepackage{graphicx}
\usepackage{subcaption}
\usepackage[para,online,flushleft]{threeparttable}
\newtheorem{theorem}{Theorem}[section]
\newtheorem{corollary}{Corollary}[theorem]

\newtheorem{assumption}{Assumption}

\newtheorem{exmp}{Example}[section]

\usepackage{longtable}
\usepackage{appendix}
\usepackage{float}
\usepackage{amssymb}
\usepackage{todonotes}
\usepackage{lipsum}
\usepackage{dsfont}
\usepackage{lmodern}
\usepackage{mathdots}
\usepackage{amsmath}
\usepackage{mathtools}
\usepackage{tikz}
\usepackage{xcolor}
\usepackage{marginnote}
\usepackage[round, sort]{natbib}
\usepackage{setspace}
\usepackage{changepage}
\usepackage{multirow}
\usepackage{authblk}
\usepackage{booktabs}
\usepackage{braket}
\usepackage{enumitem}
\usepackage[font = small, labelfont = bf]{caption}
\usepackage[a4paper, bindingoffset = 0.2in, 
left = 1in, right = 1in, top = 1in, bottom = 1in, 
footskip = .25in]{geometry}

\DeclareMathOperator{\E}{\mathbb{E}}
\DeclareMathOperator{\Cov}{\mathbb{C}ov}
\DeclareMathOperator{\Corr}{\mathbb{C}orr}

\DeclareMathOperator*{\plim}{plim}

\newcommand{\argmin}{\operatornamewithlimits{argmin}}

\numberwithin{equation}{section}
\numberwithin{figure}{section}
\usepackage[xcolor, leftbars]{changebar}
\usepackage[unicode, bookmarks = FALSE, ]{hyperref}
\hypersetup{colorlinks, citecolor=blue, filecolor=blue, linkcolor=blue, urlcolor=blue}
\usepackage{url}
\def\equationautorefname~#1\null{%
  Equation~(#1)\null
}
\usepackage{array}
\newcolumntype{L}[1]{>{\raggedright\let\newline\\\arraybackslash\hspace{0pt}}m{#1}}
\newcolumntype{C}[1]{>{\centering\let\newline\\\arraybackslash\hspace{0pt}}m{#1}}
\newcolumntype{R}[1]{>{\raggedleft\let\newline\\\arraybackslash\hspace{0pt}}m{#1}}

\graphicspath{{figures/}{../figures/}}

\usepackage{enumitem}
\usepackage{outlines}

\date{\today \\
\href{https://drive.google.com/file/d/1Xd7J4nnVzgKppNGFJMbMDc76LOA_rrhY/view?usp=sharing}{click here for the latest version}}
\author{Weifeng Jin\thanks{ Weifeng Jin: Universidad Carlos III de Madrid. \href{mailto:jweifeng@eco.uc3m.es}{jweifeng@eco.uc3m.es}} }

\title{Quantile Autoregression-Based Non-causality Testing \thanks{I would like to express my gratitude to my supervisor, Carlos Velasco, who guided me throughout this research project. I also want to thank Juan Carlos Escanciano, Miguel Delgado, Jesus Gonzalo, Juan Jose Dolado, Nazarii Salish, and Alain Hecq for their constructive comments, and also seminar and workshop participants at Universidad Carlos III de Madrid, Time Series Workshop in Zaragoza and CEBA seminar 2022.}}

\begin{document}

\maketitle

\begin{abstract}
Non-causal processes have been drawing attention recently in Macroeconomics and Finance for their ability to display nonlinear behaviors such as asymmetric dynamics, clustering volatility, and local explosiveness. In this paper, we investigate the statistical properties of empirical conditional quantiles of non-causal processes. Specifically, we show that the quantile autoregression (QAR) estimates for non-causal processes do not remain constant across different quantiles in contrast to their causal counterparts. Furthermore, we demonstrate that non-causal autoregressive processes admit nonlinear representations for conditional quantiles given past observations. Exploiting these properties, we propose three novel testing strategies of non-causality for non-Gaussian processes within the QAR framework. The tests are constructed either by verifying the constancy of the slope coefficients or by applying a misspecification test of the linear QAR model over different quantiles of the process. Some numerical experiments are included to examine the finite sample performance of the testing strategies, where we compare different specification tests for dynamic quantiles with the Kolmogorov-Smirnov constancy test. The new methodology is applied to some time series from financial markets to investigate the presence of speculative bubbles. The extension of the approach based on the specification tests to AR processes driven by innovations with heteroskedasticity is studied through simulations. The performance of QAR estimates of non-causal processes at extreme quantiles is also explored.
\noindent \bigskip



\end{abstract}
\section{Introduction}
A stationary autoregressive moving-average (ARMA) process is defined as causal concerning the specified innovation sequence if all roots of the autoregressive polynomials are outside of the unit circle, so it can be represented by an infinite sum of past innovations. However, non-causal autoregressive (AR) processes, due to their ability to display various non-linear dynamics, have been drawing increasing attention in the econometrics literature during the last two decades. The same concept has been fully exploited as a non-minimum phase in the stochastic system with applications to natural sciences. Unlike the causal AR process in the classical time series context, mixed causal and non-causal AR processes do not impose presumptions on the location of the lag polynomial roots except for the exclusion of the unit root, which allows these stationary processes to be dependent on past and future innovations at the same time. \cite{breidt2001least} showed that non-causal AR processes can capture stylized facts like clustering volatility in financial data, which usually is associated with GARCH models. The same argument is made in the paper by \cite{lanne2013testing}, where they derived a closed-form expression for the correlation of squared values in levels in the ARMA($1,1$) case.
\cite{gourieroux2017local,fries2019mixed} and \cite{cavaliere2020bootstrapping} proposed to model speculative bubbles with non-causal AR or mixed causal and non-causal AR processes generated by heavy-tailed innovations because they can display local explosive behavior. Regarding forecasting, \cite{lanne2012optimal} and \cite{hecq2021forecasting} argued that there is an accuracy gain in forecasting performance after introducing non-causality into the modeling procedure. Moreover, \cite{lanne2013autoregression} pointed out that non-causal AR processes can be an alternative to non-invertible processes for forward-looking behavior, see \cite{alessi2011non} for a comprehensive survey of empirical applications of non-invertible (non-fundamental) processes to Macroeconomics and Finance.\newline
The emerging applications of non-causal AR processes promote an interest in testing non-causality in practice, given that autocorrelation functions fail to discriminate non-causal processes from their causal counterparts. Needless to say, the non-causality check can be naturally achieved through testing classical linear hypotheses under robust estimation techniques applicable to possibly non-causal processes. These estimation methods have been developed by \cite{breid1991maximum} and \cite{lii1992approximate,lii1996maximum} through non-Gaussian maximum likelihood schemes or minimum distance estimation exploiting information contained in higher order moments/cumulants or characteristic/cumulative distribution functions of residuals, see \cite{velasco2018frequency}, \cite{velasco2022estimation}, and \cite{jin2021estimation}. With this approach, the test procedure is confined to the assumptions required for the corresponding estimates, which can be somehow stringent. Apart from that, a factorization of the coefficients is necessary for disentangling the roots, which becomes rather complicated as the order of the AR process increases. Therefore, a testing strategy before the estimation, which can potentially work as a model selection, needs investigation. Besides, the testing can serve as a detection tool for the existence of speculative bubbles in empirical time series for the ability of non-causal processes with heavy tails to exhibit local explosive behavior.\newline   
However, except for the robust estimation techniques introduced before, little has been done on the theory of testing non-causality in AR processes. Nevertheless, all the estimation techniques above deliver the same message that additional information beyond second-order moments is required to identify non-causality. Following this message, we propose some testing strategies for the non-causality of time series within the quantile autoregressions framework (QAR hereafter) (\cite{koenker2006quantile}), which allows us to make use of the complete distribution to measure dependence. Similarly, \cite{hecq2021selecting} applied QAR to the target process and entertained the sum of rescaled absolute residuals as an information criterion to select between purely causal and non-causal models. The approach of \cite{hecq2021selecting} obtains the proposed test statistic by running QAR in direct and reserved time, respectively. This approach can bring up ambiguity in the model selection when the complexity of the causality structure escalates as the order of the AR model increases. Thereby, this strategy may yield misleading results when the time series is mixed causal and non-causal.\newline
In this paper, we propose three testing procedures based on the well-developed inference for QAR estimates exploiting the non-linear characteristics of non-causal processes. Recall that the coefficients in the conditional quantile regression for the location model\footnote{The location model is any model of the form $Y = \mu(X)+ \sigma \cdot u$, where $u$ is independent of $X$. Throughout this paper, we refer to $Y_t = \mu\left(Y_{t-1}, Y_{t-2},\dots\right)+u_t$, where $\mu(\cdot)$ is a linear functional form, and $u_t$ is a sequence of $iid$ innovations, when it comes to the location model.} are quantile-invariant except for the intercept. When the process is causal, the independence of current innovation with past observations contributes to the invariant property of coefficients in the QAR model across different quantiles. However, under non-causality the true conditional quantile of a non-Gaussian AR process exhibits non-linearity in the past information. Induced from this, the best linear approximation to the true conditional quantile is expected to show varying coefficients across distinct quantiles. Using this feature, we introduce our first strategy for the objective of testing non-causality by carrying out the constancy test over the entire quantile interval. Its easy-to-implement attribute makes it a perfect candidate for a preliminary check of non-causality. The other approach to detect non-causality is achieved through a specification test of the linear functional form for the conditional quantile. With this specification-based approach, no distributional knowledge of the innovations beforehand is required, nor does the correct specification of the conditional quantile need to be spelled out.\newline
The testing strategies introduced in this paper fill a gap in the theory of testing non-causality. Apart from that, they share an appealing property of retaining power when the AR process is higher-order with a mixture of causal and non-causal representation. Like the significant advantage of quantile regression over conditional mean regression, our approach is robust to outliers, making it suitable for heavy-tailed processes commonly employed in finance. Our specification-based approach can also be tentatively extended to an AR process with conditional heteroskedasticity. The tests achieve considerable power in relatively small samples. In addition, some Monte Carlo simulation results suggest that the estimates in linear quantile models approach the true parameters for non-causal processes as the quantile estimand approaches to either extremum ($0$ or $1$), which can be a potential viewpoint to investigate the nonlinear features of non-causal processes in the future research.\newline    
The rest of the paper is organized as follows. The second section introduces mixed causal and non-causal Autoregressions and some of their statistical properties. Section 3 investigates the non-causality testing within the QAR framework. Section 4 discusses the finite sample performance of our proposed testing procedures through Monte Carlo Simulations. Section 5 illustrates the tests using financial data with the possible existence of speculative bubbles. Section 6 closes the paper with conclusions and some extensions. 
\section{Mixed Causal and Non-causal Autoregressions} 
\begin{figure}[h]
    \centering
    \includegraphics[width=\textwidth]{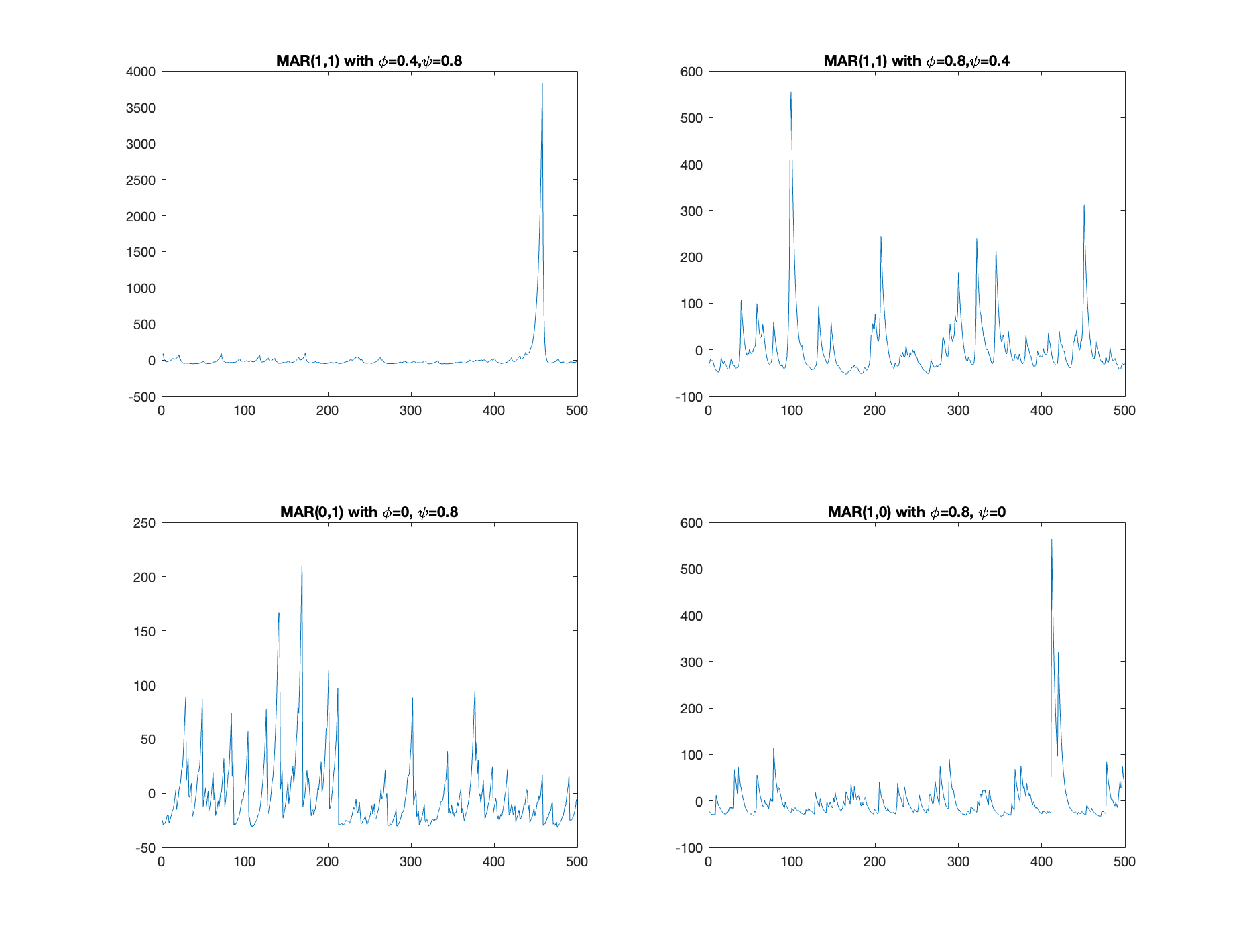}
    \caption{Trajectories of MAR($1,1$) processes with different parameters $(\phi, \psi)$, T=500: the upper panel exhibits two MAR($1,1$) processes with different parameters on the causal/non-causal components; the lower panel exhibits a purely non-causal AR(1) process (left) and a purely causal AR(1) process (right) when one of the parameters degenerates to zero.}
    \label{trajectory MAR}
\end{figure}

In the context of classical time series analysis, it is customary to restrict attention to the causal representation of autoregressive processes while modeling stationary univariate time series. The reason is mentioned in \cite{brockwell2009time}. Every non-causal autoregressive process is a stationary solution to a future-independent autoregressive process with explosive roots. It provides the same second-order structure as its causal counterpart. However, to feature higher-order dynamics, a general framework of autoregressive processes named mixed causal and non-causal autoregressions (MAR($r,s$) hereafter) is proposed where temporal dependence in both past and future is introduced in the processes, defined by
\begin{equation}\label{mixed autoregression}
\phi(L)\psi(L^{-1})Y_t = u_t
\end{equation}
where $\phi(L)=1-\phi_1L-\phi_2L^2-\dots-\phi_r L^r$ and $\psi(L^{-1})=1-\psi_1L^{-1}-\psi_2L^{-2}-\dots-\psi_{s} L^{-s}$ are polynomials with backward and forward operators, respectively. $\{u_t\}_{t\in \mathbb{Z}}$ is a sequence of independent identically distributed $( iid )$ innovations with zero mean. $p=r+s$ is the total order encompassing both causal and non-causal polynomials. An equivalent expression of equation (\ref{mixed autoregression}) in moving average can be given by
\begin{equation}\label{double MA}
Y_t =\phi(L)^{-1}\psi(L^{-1})^{-1}u_t = \sum_{j=-\infty}^{\infty}\rho_j u_{t-j},
\end{equation}
where $Y_t$ may depend on both future and past innovations. The stationarity of $Y_t$ is assured by the absolute summability of $\rho_j$, $\sum_{j=-\infty}^{\infty}\left\vert \rho_j\right\vert < \infty $ and $\E\left\vert u_t \right\vert^{1+\delta} < \infty$ for $\delta >0$. The former condition is guaranteed once the roots of both polynomials $\phi(z)$ and $\psi(z)$ are confined to locate outside the unit circle.\newline
When $\psi(z)= 1$ for all $z$, the process is reduced to a purely causal process MAR($r,0$) like in conventional studies. While if $\phi(z)=1$ for all $z$, the process becomes purely non-causal. Below, we attach several examples to illustrate the statistical properties of a general autoregressive process MAR($r,s$).
\begin{exmp}[Second-order Property]\label{second-order property}
Define a purely non-causal MAR($0,1$) sequence $(1-\psi L^{-1})Y_t=u_t$ starting from an $iid$ sequence of innovations $u_t$ of zero-mean and finite variance $\sigma^2$. The autocovariance function of $Y_t$ is provided by
\begin{equation*}
    \begin{split}
        \Cov\left(Y_{t+h}, Y_{t}\right) = & \Cov\left(\sum_{j=0}^{\infty}\psi^{j}u_{t+h+j},\sum_{j=0}^{\infty}\psi^{j}u_{t+j}\right) \\
        = & \sum_{j=0}^{\infty}\Cov\left(\psi^{j}u_{t+h+j},\psi^{j+h}u_{t+h+j}\right) \\
        =&\frac{\psi^h}{1-\psi^2}\sigma^2 \text{ for }h=0,1,2,\dots
    \end{split}
\end{equation*}
It is worth noting, after some simple calculation, that $\{Y_t\}_{t \in \mathbb{Z}}$ shares the same autocovariance function as the MAR($1,0$) sequence $\{\tilde{Y}_t\}_{t \in \mathbb{Z}}$ defined by $(1-\psi L)\tilde{Y}_t=u_t$, which is its causal counterpart. A general conclusion can be drawn for MAR($r,s$) through the autocovariance generating function (ACGF). The ACGF of a stationary autoregression is given by 
\begin{equation*}
\begin{split}
   G(L) & =\frac{1}{\left\vert \phi(L)\psi(L^{-1})\right\vert^2}\sigma^2  = \frac{1}{\phi(L)\phi(L^{-1})\psi(L^{-1})\psi(L)}\sigma^2 \\
   & = \frac{1}{\phi(L)\psi(L)\phi(L^{-1})\psi(L^{-1})}\sigma^2\\
   & = \frac{1}{\left\vert\phi(L)\psi(L)\right\vert^2}\sigma^2,
   \end{split}
\end{equation*}
from where it is clear that MAR($r,s$) takes the identical form of ACGF with MAR($r',s'$) as long as the total order $r+s=r'+s'$ is satisfied and the roots to all polynomials match.
\end{exmp}
This second-order property explains the failure to distinguish non-causal processes from causal processes based on the ACF. Moreover, it implies that non-Gaussianity of innovations is required for identifying non-causal processes since second-order properties are sufficient to characterize a Gaussian probabilistic structure but not to other distributions.\newline
\begin{exmp}[Local Explosiveness]\label{local explosiveness}
Define a MAR($1,1$) process by
\begin{equation*}
(1-\phi L)(1-\psi L^{-1})Y_t = u_t, \text{ where } u_t \sim Lognormal(0,2)-\exp(2) \end{equation*}
\end{exmp}
A MAR($r,s$) process with heavy-tailed innovations can generate multiple phases of local explosiveness, which can be employed to model speculative bubbles. A detailed investigation of probabilistic properties of a MAR process driven by $\alpha$-stable non-Gaussian innovations is provided by \cite{fries2019mixed}, where they show the properties of the marginal and conditional distributions of a stable MAR($r,s$). But in a general situation, there is no closed-form solution to either the marginal or the conditional distribution of a non-causal AR process. Here we illustrate the potential applications of MAR($r,s$) processes in modeling speculative bubbles with some simulated trajectories. In this example, we plot four distinct scenarios by varying the parameters of both causal and non-causal components of the MAR($1,1$) process with log-normal distributed innovations\footnote{A log-normal distribution is a heavy-tailed continuous distribution defined on the positive domain, whose density function is characterized by the location parameter $\mu$ and scale parameter $\sigma$. The mean and variance are represented by $exp(\mu+\sigma^2/2)$ and $\left(exp(\sigma^2)-1\right)exp\left(2\mu+\sigma^2\right)$, respectively. In this example, we employ centered log-normal distribution to be compatible with our setup of mean zero.}. Generally, with non-causality, the processes can mimic bubbles by repetitive phases of upward trends followed by a sharp drop; see the lower panel in Figure \ref{trajectory MAR}. The upper panel in the same figure indicates more complicated dynamics can be generated by incorporating more causal/non-causal components in the data-generating process regarding the number and magnitude of bubbles.
\begin{exmp}[Conditional heteroskedasticity]\label{ex3}
Here we exemplify the applicability of MAR($r,s$) in characterizing clustering volatility as an alternative to ARCH or other stochastic volatility models with a simple MAR($1,1$) process. This argument is originally made by \cite{breidt2001least} in an empirical application to New Zealand/US exchange rate. \cite{lanne2013testing} elaborate on it by deriving explicitly the expression of the autocorrelation of $Y^2_t$ in an all-pass model of order 1. However, no formal justification for the higher-order dependence structure has been made on general non-causal processes. Nevertheless, we present the following example that a non-causal process can exhibit higher-order dependence.\newline
Continued with data generating process MAR($1,1$), it can be reexpressed by an AR(2) with the $\frac{1-\psi L}{1-\psi L^{-1}}$ filter on the innovations.
\begin{equation*}
    \begin{split}
        & (1-\phi L)(1-\psi L^{-1})Y_t = u_t \quad{} u_t \sim IID(0,\sigma^2)\\
        \Longleftrightarrow & 
        (1-\phi L)(1-\psi L)Y_t = \tilde{u}_t \\
         & \text{ with }\tilde{u}_t = \frac{1-\psi L}{1-\psi L^{-1}}u_t =\sum_{j=-\infty}^{\infty}\rho_ju_{t+j} \\
         & \text{ with } \rho_j =\left\{ \begin{array}{ll}
             0 & \text{ if } j <-1 \\
             -\psi & \text{ if } j=-1\\
             \psi^j-\psi^{j+2} & \text{ if } j\geq 0.
         \end{array} \right.
    \end{split}
\end{equation*}
\end{exmp}
The $\frac{1-\psi L}{1-\psi L^{-1}}$ filter introduces higher-order dependence to an $iid$ innovation sequence\footnote{Actually, $\tilde{u}_t$ is an all-pass time series model, where all roots of the autoregressive polynomial are reciprocals of roots in the moving average polynomial and vice versa.} preserving its uncorrelatedness. By simple algebra, we can obtain the formula of the autocorrelation function of $\tilde{u}^2_t$,
\begin{equation*}
    \Corr\left(\tilde{u}^2_t, \tilde{u}^2_{t+h}\right) = \frac{\kappa_4(u_t)\left(\sum_{j=-\infty}^{\infty}\rho^2_j\rho^2_{j+h}\right)+2\sigma^4\left(\sum_{j=-\infty}^{\infty}\rho_j\rho_{j+h}\right)^2 }{\kappa_4(u_t)\left(\sum_{j=-\infty}^{\infty}\rho^4_j\right)+2\sigma^4\left(\sum_{j=-\infty}^{\infty}\rho^2_j\right)^2}
\end{equation*}
which, in general, does not vanish\footnote{ $\kappa_4(X)$ is the fourth cumulant of the random variable $X$, defined by $\kappa_4(X)= \E\left(\left(X-\E(X)\right)^4\right)-3\E^2\left(\E\left(X-\E(X)\right)^2\right)$. }. This property demonstrates the capability of non-causal processes to exhibit volatility clustering, which is commonly observed in financial data. The higher-order dependence analysis of $\tilde{u}_t$ is relegated to Appendix \ref{higher order dependence}, indicating the possibility of accommodating higher-order nonlinear dynamics with a general MAR($r,s$) model.\newline
As shown in the preceding examples, MAR($r,s$) models can display various nonlinear characteristics with a linear process generation scheme. This deviation from "linearity" can be employed as a crucial feature to detect non-causality in the linear time series. A fundamental result is formalized by \cite{rosenblatt2000gaussian} on the best one-step predictor in the mean square for a general AR process, where he demonstrates that the conditional expectation must be nonlinear in the past if there is non-causality involved in the non-Gaussian AR processes with finite variance. This statement provides us with the critical theoretical result where our tests for non-causality are grounded, which will be elaborated on in the next section. The extension to a general VARMA process framework has been developed by \cite{chen2017testing} and applied to a test for non-invertibility. Afterward, \cite{fries2019mixed} consider the case when the MAR($1,s$) process is driven by symmetric $\alpha$-stable innovations with infinite variance. They surprisingly find that the conditional expectation can be explicitly expressed by a linear function of the past information, in contrast to a MAR($r,s$) model with finite variance. However, no study has yet been done on the distributional characterization of a MAR($r,s$) process due to no closed-form solution. Following a simulation-based approach, we perform a preliminary analysis of the conditional density function of a process given the past observations, see Appendix \ref{conditional density function}. In short, in the presence of non-causality, the shape of the conditional distribution of the process, say $f(Y_t|Y_{t-1}=y)$, is $y$-dependent. Still, the dependence pattern is hard to characterize since it varies across different distributions.  
\section{QAR-based Non-causality Tests}
\subsection{Benchmark model: MAR($r,s$)}
\begin{figure}[h]
    \centering
    \begin{subfigure}[h]{0.4\textwidth}
    \centering
    \includegraphics[width=\textwidth]{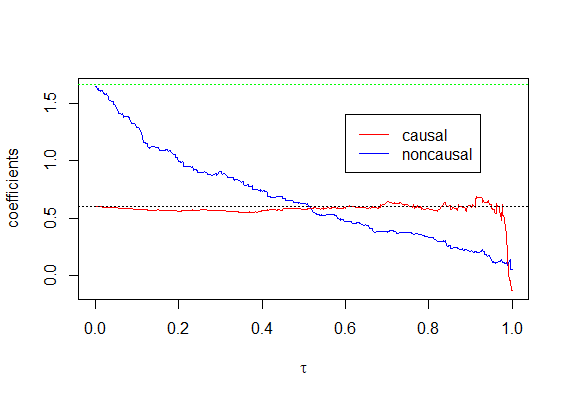}
    \caption{QAR(1) estimates of the slope of $Y_{t-1}$ over (0,1): exponential distribution}
    \label{qarex}
    \end{subfigure}
\begin{subfigure}[h]{0.4\textwidth}
    \centering
    \includegraphics[width=\textwidth]{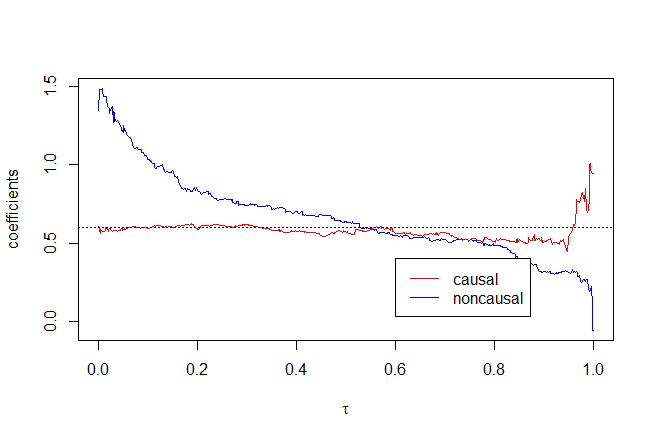}
    \caption{QAR(1) estimates of the slope of $Y_{t-1}$ over (0,1): chi-square distribution
    }
    \label{qarchi}
    \end{subfigure}
    \caption{QAR($1$) estimates of a pair of AR(1) processes: one is causal, and the other is non-causal. The left part of the figure is applied to the processes generated by exponential innovations, and the right part is to the processes generated by chi-square innovations. The true parameter is $0.6(1/0.6)$ for the causal(non-causal) case. }
    \label{qar1}
\end{figure}

In this section, we formalize the test procedures for non-causality. The null hypothesis of interest is $Y_t$ being a causal process, i.e.
\begin{equation}
 \mathbb{H}_0: \psi_1=\dots=\psi_s=0  \text{ in } (\ref{mixed autoregression})  
\end{equation}
against the alternative hypothesis denoted by $\mathbb{H}_A$, which is $\{Y_t\}$ is non-causal,
\begin{equation}
    \mathbb{H}_A: \psi_k \neq 0 \text{ for some } k \in \{1,2,\dots, s\} \text{ in } (\ref{mixed autoregression}).   
\end{equation}
A primitive strategy is to take it as a joint significance test for the coefficients of leads $\{Y_{t+j}\}_{j=1,2,\dots,s}$. This approach would call for the identification of models and robust inference for the estimates under both hypotheses. In this paper, we propose a test for $\mathbb{H}_0$ employing the linearity property of $\{Y_t\}$ based on the QAR, which is easy to implement empirically and does not require consistent estimates for general autoregressive progresses.\newline  
Denoting the information set generated by the observations up to period $t$ by $I_{t}=\sigma\left(Y_{t}, Y_{t-1},...\right)$ and the $\tau-$th quantile of $Y_t$ conditional on the past by $Q_{Y_t}\left(\tau|I_{t-1}\right)$, the following result on the linearity of $\{Y_t\}$ justifies our approach. 
\begin{assumption}\label{ASS1}
Let $\{u_t\}_{t\in \mathcal{Z}}$ be a non-Gaussian $iid$ sequence with $(k+1)$th order moment finite and $(k+1)$th cumulant nonzero for some $k \geq 2$. 
\end{assumption}
\begin{theorem}\label{linearity}
Under Assumption \ref{ASS1}, a stationary MAR($r,s$) process $\{Y_t\}$ has a non-degenerated non-causal component, i.e., $s \neq 0$ if and only if there exists $Q_{Y_t}\left(\tau|I_{t-1}\right)$ nonlinear in $\{Y_{t-j}\}_{j\geq 1}$ for at least one $\tau \in (0,1)$.  
\end{theorem}
As discussed in Example \ref{second-order property}, there is no meaning to discuss non-causality in a Gaussian structure as there is always an equivalent causal representation of a Gaussian AR process for its non-causal counterpart. The finiteness of the moment condition and  nonzero cumulants are also required in \cite{rosenblatt2000gaussian}. Theorem \ref{linearity} is deduced from the nonlinearity of the best predictor of $Y_t$ in the mean square criterion for non-causal processes. The nonlinearity in the conditional mean implies dependence in the conditional distribution beyond the linear correlation. Note that the theorem does not point out the quantile(s) where this nonlinear relationship occurs, nor the manners in which this nonlinearity is expressed. Another remark is that the information set considered in the theorem can be replaced by $\sigma$-field generated by $Y_{t-1}$ up to $Y_{t-p}$ due to the Markovian property of MAR($r,s$), which avoids the infinite-dimensional issue arising from $I_{t-1}$. The total number of the lags and leads of $Y_t$ included to explain the conditional quantile of MAR($r,s$) processes, $p,$ is determined by the partial autocorrelation function (PACF) of $Y_t$.\newline
This theorem prompts us to adopt QAR as a medium to detect non-causality. QAR is a comprehensive analysis tool in the time series context, providing robust statistical analyses against outliers in the measurement of the response variable, which has proven to be rather prevailing in recent decades. Given a MAR($r,s$) process of order $p$ defined by (\ref{mixed autoregression}), if the non-causal component degenerates to 1, i.e., $s=0, r=p$, the conditional quantile of $Y_t$ can be expressed by 
\begin{equation}\label{qarmodel}
\begin{split}
Q_{Y_t}\left(\tau \mid I_{t-1}\right)= & Q_{u_t}\left(\tau\right)+ \phi_1 Y_{t-1}+\phi_2 Y_{t-2}+\dots+\phi_p Y_{t-p}\\
= & \theta_0(\tau)+\theta_1(\tau)Y_{t-1}+\theta_2(\tau)Y_{t-2}+\dots+\theta_p(\tau)Y_{t-p}\\
= & \boldsymbol{X}'_{t}\boldsymbol{\theta}(\tau) \quad{} \forall \tau \in (0,1),
\end{split}
\end{equation}
where $Q_{u_t}(\tau)$ denotes the $\tau$-quantile of innovations $u_t$, and $\phi_j$'s are the coefficients of corresponding $Y_{t-j}$ in the polynomial expansion of (\ref{mixed autoregression}). Therefore, after imposing pure causality, the conditional quantile of $Y_t$ can be fully characterized by a linear function of past observations, $\boldsymbol{X}'_t\boldsymbol{\theta}(\tau)$ with $\boldsymbol{X}_t=\left(1, Y_{t-1}, Y_{t-2},\dots, Y_{t-p}\right)^{'}$ and $\boldsymbol{\theta}(\tau)=\left(\theta_0(\tau),\dots,\theta_p(\tau)\right)^{'}$. The QAR estimates of coefficients $\boldsymbol{\theta}(\tau)$ in this linear quantile model can be obtained by minimizing the following problem,
\begin{equation}\label{lossfunctionqar0}
\hat{\boldsymbol{\theta}}(\tau) = \argmin_{\boldsymbol{\theta}\in \mathbb{R}^{p+1}}\sum_{t=1}^{T}\rho_{\tau}(Y_t-\boldsymbol{X}'_t\boldsymbol{\theta}),
\end{equation}
with the check function $\rho_{\tau}(u)=u\left(\tau-\mathbb{I}(u<0)\right)$. The asymptotic properties of linear QAR estimates were first established by \cite{koenker2006quantile}. A brief review of QAR estimates (\ref{lossfunctionqar0}) can be found in Appendix \ref{qarasymptotics}. However, if $Y_t$ has a non-degenerated non-causal component, the linear dynamic model (\ref{qarmodel}) for its conditional quantile is misspecified for at least one $\tau \in (0,1)$, following Theorem \ref{linearity}.\newline
Within the linear QAR framework, \cite{hecq2021selecting} consider a statistic aggregating the information of the residuals over quantiles, which they employ as a model selection criterion between purely causal AR models and purely non-causal AR models. Given that the calculation of residuals is done either by running QAR with direct or reversed time, this methodology may provide misleading conclusions regarding MAR($r,s$) in presence of causality and non-causality at the same time. By contrast, our approaches address more the correctness of the linear specification of the conditional quantile through the QAR. For non-causal autoregressive processes, no closed-form of the nonlinear conditional quantile of $Y_t$ is required. Consequently, our approach is robust to the general MAR($r,s$) setting. Before we carry out the tests for non-causality, the following assumptions are imposed in the QAR framework. 
\begin{assumption}\label{ASS2}
The distribution function of innovations $u_t$, $F(u)$ admits a continuous density function $f(u)$ away from zero on the domain $\mathcal{U}=\{u: 0<F(u)<1\}$. 
\end{assumption}
\begin{assumption}\label{ASS3}
Denote the family of conditional distribution $\{P\left(Y_t < y|\boldsymbol{X}_t=x \right), y \in \mathbb{R}, x \in \mathbb{R}^{r+s}\}$ as $F_x(y)$ and its Lebesgue density as $f_x(y)$, that is uniformly bounded on the space of $y\times x \subseteq \mathbb{R}\times \mathbb{R}^{r+s}$ and uniformly continuous.
\end{assumption}
\begin{corollary}\label{col2}
Under Assumptions \ref{ASS1}-\ref{ASS3} and null hypothesis $\mathbb{H}_0$, the coefficients except for the intercept of the conditional quantile are constant across different quantiles in $(0,1)$.
\end{corollary}
Corollary \ref{col2} is a consequence of the independence between $u_t$ and $Y_{t-j}$ for $j=1,2,\dots,p$ when the MAR($r,s$) is purely causal. As shown in the equation (\ref{qarmodel}), the slope coefficients $\{\theta_j(\tau)\}_{j=1,\dots,p}$ are constant over $\tau \in (0,1)$ and uniquely determined by the expansion of autoregressive polynomials of $Y_t$. Instead, the performance of the QAR estimates is more complicated in the mixed causal and non-causal autoregressive processes since the linear model is under misspecification. \cite{angrist2006quantile} demonstrated in their paper that quantile regression is essentially an approximation to the true conditional quantile function in a weighted mean squared criterion, with weights associated with the densities of $Y_t$ ranging from the linear approximation (\ref{qarmodel}) to the true conditional quantile $Q_{Y_t}(\tau|Y_{t-1},\dots, Y_{t-p})$. Their statement presents a rough idea of how well the linear function fits the true conditional quantile.\newline
\paragraph{Constancy Test}
Our first approach for detecting 
non-causality comes along with the constancy test of QAR coefficients over the entire quantile range. As stated in Corollary \ref{col2}, if the process is causal, the constancy should hold for all $\theta_j(\tau)$ for all $j=1,2,...,p$ and $\tau \in (0,1)$. Whereas for a non-causal process, the estimated coefficients of $Y_{t-j}$ in the quantile regression may vary across different quantiles much likely for the following intuitions: i) non-causal processes display highly nonlinear dynamics, one of which is asymmetric dynamics; ii) linear quantile model is misspecified; iii) the conditional distribution of $Y_t$ varies both in the location and shape at different values of past observations. For instance, Figure \ref{qar1} depicts the dynamic performance of QAR(1) estimates of the slope parameter across the quantiles for a non-Gaussian autoregressive process, including causal and non-causal cases in different colors. The QAR(1) estimates in the non-causal processes (solid blue lines) exhibit a trendy pattern over the quantile domain. However, since a general solution for the true conditional quantile function of $Y_t$ is infeasible, we do not attempt to provide a detailed characterization of the varying coefficient property in the non-causal situation. The test for non-causality based on the constancy test can only be used to check the necessary condition of AR processes being causal. Nevertheless, the accessibility and straightforwardness of this method make it a touchstone for non-causality testing in practice.\newline
\newline
To implement the constancy test of coefficients under the QAR framework, we consider the approach developed by \cite{koenker2006quantile}, where the hypothesis is formulated in the manner analog to the classical linear hypothesis: $$\mathbb{H}^{1}_0: \boldsymbol{R\theta}(\tau) = \boldsymbol{\phi} \text{ with } \boldsymbol{R}= \left(\boldsymbol{0}_{p\times 1}\vdots \boldsymbol{I}_{p}\right) \text{ for all } \tau \in (0,1)$$
with the unknown $\tau$-invariant parameter vector $\boldsymbol{\phi} = \left(\phi_1,\phi_2,\dots,\phi_p\right)'$ which needs to be estimated\footnote{For purely causal processes, the constant vector consists of the parameters in the autoregressive polynomials, i.e., $\phi_1, \phi_2,\dots,\phi_p$ in (\ref{qarmodel}) }. Naturally, the naive test for this hypothesis is constructed on the quantile process
\begin{equation*}
    V_{T}\left(\tau\right) = \sqrt{T}\left[\boldsymbol{R}\hat{\Sigma}^{-1}_1\hat{\Sigma}_0\hat{\Sigma}^{-1}_{1}\boldsymbol{R}'\right]^{-1/2}\left(\boldsymbol{R}\boldsymbol{\hat{\theta}}(\tau)-\boldsymbol{\hat{\boldsymbol{\phi}}}\right)
\end{equation*}
and the Kolmogorov-Smirnov (KS) type of test statistic is adopted for the interest of testing a compact set of quantiles
\begin{equation*}
    KSV_{T} = \sup_{\tau \in \mathcal{T} \subset (0,1)}V_{T}\left(\tau\right), \text{ where } \mathcal{T} \text{ is a compact interval},
\end{equation*}
where $\hat{\boldsymbol{\theta}}(\tau)$ is the linear QAR estimate, and $\hat{\Sigma}_1, \hat{\Sigma}_0$ are the corresponding estimates of the asymptotic variance, see appendix \ref{qar}. $\hat{\boldsymbol{\phi}}$ is a $\sqrt{T}$ consistent estimator of $\boldsymbol{\phi}$ and a simple choice is the QAR estimator $\hat{\boldsymbol{\theta}}(\tau^{*})$ at any $\tau^{*} \in \mathcal{T}$.\footnote{Another appropriate choice is the estimator from the ordinary least square of $\phi_{j}, j=1,2,\dots,p$ in (\ref{mixed autoregression}).} A closed interval $[\epsilon_1,1-\epsilon_2]$ with trivial numbers $\epsilon_1, \epsilon_2$ is proposed for $\mathcal{T}$ to avoid missing much information from the entire quantile interval $(0,1)$.\newline
Under the hypothesis of constancy $\mathbb{H}_0^1$,  
$$
V_{T}\left(\tau\right) \Longrightarrow \boldsymbol{B}_{p}(\tau)-f\left(F^{-1}(\tau)\right)\left[\boldsymbol{R}\Sigma_0^{-1}\boldsymbol{R}'\right]^{-1/2}\boldsymbol{Z}
$$
where $\boldsymbol{B}_{p}(\tau)$ is a $p$-dimensional standard Brownian bridge and $\boldsymbol{Z}=\plim_{T \rightarrow \infty} \sqrt{T}\left(\hat{\boldsymbol{\phi}}-\boldsymbol{\phi}\right)$ is a drift brought up by the estimation of the nuisance parameter $\boldsymbol{\phi}$. To annihilate this non-trivial effect,  a martingale transformation $\mathcal{K}$ was introduced into $V_{T}\left(\tau\right)$ to retrieve the distribution-free merit of the KS test. Denote the derivative of the density function by $\dot{f}$ and define 
\begin{equation*}
    \begin{split}
        g(x) & = \left(1, \left(\dot{f}\left(F^{-1}(x)\right)/f\left(F^{-1}(x)\right)\right)\right)' \text{ and }\\ 
        C(z) & = \int_{z}^{1}g(x)g(x)'dx, 
    \end{split}
\end{equation*}
and the martingale transformation on the process $V_{T}\left(\tau\right)$ is constructed as follows
\begin{equation*}
\begin{split}
    \tilde{V}_{T}\left(\tau\right) = & \mathcal{K}V_{T}\left(\tau\right) \\
    = & V_{T}\left(\tau\right) - \int_{0}^{\tau}\left[g'_{T}(x)C^{-1}_{T}(x)\int_{x}^{1}g(s)dV_{T}\left(s\right)\right]dx,
\end{split}    
\end{equation*}
where $g_{T}(x)$ and $C_{T}(x)$ are uniformly consistent estimators of $g(x)$ and $C(x)$ in the considered domain, respectively. The proposed KS-type norm on the transformed process becomes
\begin{equation*}
    KS\tilde{V}_{T} = \sup_{\tau \in \mathcal{T}}\left\Vert\tilde{V}_{T}\left(\tau\right)\right\Vert.
\end{equation*}
\begin{corollary}[Constancy test for non-causality]\label{theoremcons}
Under Assumptions \ref{ASS1}-\ref{ASS3} and the causality hypothesis $\mathbb{H}_0$,
\begin{equation*}
    \begin{split}
       \tilde{V}_{T}\left(\tau\right) & \Rightarrow \boldsymbol{W}_{p}(\tau) \\
       KS\tilde{V}_{T} &  \Rightarrow \sup_{\tau \in \mathcal{T}}\left\Vert\boldsymbol{W}_{p}(\tau) \right\Vert,
    \end{split}
\end{equation*}
where $\boldsymbol{W}_p(\tau)$ represents a $p$-dimensional standard Brownian motion. 
\end{corollary}
The related discussion on the estimation of density and score functions is given in \cite{koenker2002inference}, providing suggestions on the choice of bandwidth in detail. In the command \texttt{KhmaladzeTest} implemented in R studio, Hall/Sheather bandwidth (\cite{hall1988distribution}) for sparsity estimation is set as default. The critical values are obtained through approximating $\boldsymbol{W}_p(\tau)$ by a Gaussian random walk, and the corresponding values at different significance levels can be found in tables in the Appendix of \cite{koenker2002inference}. One remark on the quantile interval in the proposition, typically a symmetric interval $\left[\epsilon, 1-\epsilon\right]$ trimmed by a small number $\epsilon $ close to 0 is considered for simplicity in practice. Monte Carlo experiments have evidenced that an appropriate trimming in the entire quantile interval alleviates the over-rejection of the null hypothesis ascribable to the instability of estimation at extremal quantiles without sacrificing the power.\newline
\paragraph{Specification test-based approach}
Another direction to test non-causality is based on Theorem \ref{linearity}, where non-causality in the linear processes is translated into the misspecification of conditional quantiles of non/causal processes by linear dynamic quantile models (\ref{qarmodel}). Equivalently, we aim to test 
\begin{equation}\label{specificationtest}
    \E\left(\Psi_{\tau}\left(Y_t-\boldsymbol{X}'_t\boldsymbol{\theta}_0\right)|Y_{t-1},\dots,Y_{t-p}\right)=0 \text{ a.s. for some } \boldsymbol{\theta}_0 \in \mathcal{B} \text{ and }\forall \tau \in \Upsilon \subset (0,1),  
\end{equation}
where $\Psi_{\tau}(\cdot)= \mathbb{I}\left(\cdot \leq 0\right)-\tau$, and $\mathcal{B}$ is a family of uniformly bounded functions from $\Upsilon \subset (0,1)$ to $\Theta \subset \mathbb{R}^{p+1}$.\footnote{In our case, actually only $\theta_0(\tau)$ is required to be a uniformly bounded function from $\Upsilon \subset (0,1)$ to $\Theta \subset \mathbb{R},$ and the rest $\theta_j(\tau)$ are mapped to a constant for $j=1,2,\dots,p. \ \forall \tau \in (0,1)$.} Both $\Upsilon$ and $\Theta$ are compact sets. \cite{escanciano2010specification} (hereafter EV) characterize this restriction by unconditional moments
\begin{equation}\label{ev}
    \E\left(\Psi_{\tau}\left(Y_t-\boldsymbol{X}'_t\boldsymbol{\theta}_0\right)\exp\left(ix'\boldsymbol{X}_t\right)\right)=0, \quad{} \forall x \in \mathbb{R}^{p+1}, \text{ for some } \boldsymbol{\theta}_0 \in \mathcal{B} \text{ and } \forall \tau \in \Upsilon \subset (0,1),
\end{equation}
where $i=\sqrt{-1}$.
Following the strategy of the EV test, we consider the statistic based on the residual processes indexed by quantiles $\tau$, $\boldsymbol{\theta} \in \mathcal{B}$ and $x \in \mathbb{R}^{p+1}$
\begin{equation}\label{evtest}
    R^{EV}_{T}\left(x,\tau; \hat{\boldsymbol{\theta}}\right) = T^{-1/2}\sum_{t=1}^{T}\Psi_{\tau}\left(Y_t-\boldsymbol{X}'_t\hat{\boldsymbol{\theta}}\right)\exp\left(ix'\boldsymbol{X}_t\right) 
\end{equation}
with true parameters $\boldsymbol{\theta}_0$ replaced by QAR estimates $\hat{\boldsymbol{\theta}}$ from a given sample $\left(Y_t, \boldsymbol{X}^{'}_t\right)_{t=1,2,\dots,T}$. $\boldsymbol{X}_t$ is composed by a constant and the lags of $Y_t$ up to order $p$. Theoretically, $T^{-1/2}R^{EV}_{T}$ approaches to zero at a certain rate when $T$ goes to infinity for any $x \in \mathbb{R}^{p+1} $ if $\boldsymbol{X}'_t\boldsymbol{\theta}_0(\tau)$ is the correct specification for $Q_{Y_t}(\tau|I_t)$ and $\hat{\boldsymbol{\theta}}(\tau)$ is a $\sqrt{T}$-consistent estimator of $\boldsymbol{\theta}_0(\tau)$ for $\tau \in \Upsilon$. Thus, the distance between this statistic (\ref{evtest}) and zero naturally turns into a measure of the deviation of $\boldsymbol{X}'_t\boldsymbol{\theta}_0(\tau)$ from the true $Q_{Y_t}(\tau|I_t)$.
The suggested Cram\'er-von Mises (CvM) norm on (\ref{evtest}) is defined by
\begin{equation}\label{ev_cvm}
\begin{split}
    & \int_{\tau \in \Upsilon, x\in \mathbb{R}^{p+1}}\left\vert R^{EV}_T\left(x,\tau;\hat{\boldsymbol{\theta}}\right)\right\vert^2d\Phi(x)dW(\tau),
    \end{split}
\end{equation}
where $\Phi(x)$ and $W(\tau)$ are weighting functions defined on $\mathbb{R}^{p+1}$ and $\Upsilon$, respectively with positive derivative in the corresponding domain. This CvM norm 
permits us to consider infinite model specifications for conditional quantiles at all $\tau$'s of interest. Other possible options of norms aggregating the information over the quantiles and $x$, for instance, Kolmogorov-type, are also applicable here. The following proposition presents the asymptotic behavior of the EV test applied for testing non-causality.
\begin{corollary}[EV Test for non-causality]\label{theoremev}
Under $\mathbb{H}_0$ and Assumptions \ref{ASS1}-\ref{ASS3}, let $\E\left(\boldsymbol{X}_t\boldsymbol{X}'_t\right)$ be nonsingular in a neighborhood of $\boldsymbol{\theta}(\tau)=\boldsymbol{\theta}_0(\tau)$ for all $\tau \in \Upsilon$,
\begin{equation*}
    R^{EV}_{T}\left(x,\tau; \hat{\boldsymbol{\theta}}\right) \Longrightarrow \tilde{R}^{EV}_{\infty}\left(x,\tau\right),
\end{equation*}
and 
\begin{equation*}
    \int_{\tau \in \Upsilon, x\in \mathbb{R}^{p+1}}\left\vert R^{EV}_T\left(x,\tau; \hat{\boldsymbol{\theta}}\right)\right\vert^2d\Phi(x)dW(\tau) \longrightarrow_{d} \int_{\tau \in \Upsilon, x\in \mathbb{R}^{p+1}}\left\vert \tilde{R}^{EV}_{\infty}\left(x,\tau\right)\right\vert^2d\Phi(x)dW(\tau)
\end{equation*}
where $\tilde{R}^{EV}_{\infty} = R_{\infty}-\Delta R$. $R_{\infty}$ is a Gaussian process with mean zero and covariance function defined by 
$$
\Cov\left(x_1,x_2;\tau_1,\tau_2\right)= \left(\tau_1 \wedge \tau_2-\tau_1\tau_2\right)\E\left(\exp\left(i(x_1-x_2)'\boldsymbol{X}_{0}\right)\right)
$$
and the drift $\Delta R$ is introduced due to the asymptotic effect from estimation error of $\hat{\boldsymbol{\theta}}$,
$$
\Delta R(x,\tau)= G'(x,\boldsymbol{\theta}_0(\tau))Q(\tau), 
$$
where $G(x,\boldsymbol{\theta}_0(\tau)) =\E\left[\boldsymbol{X}_tf\left(Q_{u_t}(\tau)\right)\exp(ix'\boldsymbol{X}_t)\right]$ and $Q(\cdot)$ is $\Sigma^{-1/2}_0\boldsymbol{B}_{p+1}/f\left[F^{-1}(\cdot)\right].$ $\boldsymbol{B}_{p+1}$ is a $(p+1)$-dimensional standard Brownian bridge. $f$ and $F$ are the density and cumulative distribution functions of the innovation $u_t$, respectively. $\Sigma_0= \E\left(\boldsymbol{X}_t\boldsymbol{X}'_t\right)$. 
\end{corollary}
The result immediately follows from \cite{escanciano2010specification}, where they develop the result for a general class of quantile estimates covering various linear and nonlinear models of interest with corresponding assumptions. Those conditions are satisfied under Assumptions \ref{ASS1}-\ref{ASS3} in the context of MAR($r,s$) processes under the null hypothesis. The limiting distribution of the test statistic is no longer distribution-free due to the estimation of nuisance parameters. Hence, a subsampling method is proposed to approximate the critical value. The operation for calculating the residual process (\ref{evtest}) and the test statistic (\ref{ev_cvm}) is applied to a given subsample $\left(Y_t,\dots, Y_{{t+b}}\right)$ of size $b$, denoted by $R^{EV}_{b}(x,\tau;\hat{\boldsymbol{\theta}}_{b,t})$ and $\Gamma\left(R^{EV}_{b,t}\right)$ respectively, and repeated for $t=1,2,\dots, T-b+1$. The cdf of the limiting distribution of the proposed statistic is approximated by the empirical cdf across resamples, i.e., $$\hat{P}\left[\Gamma\left(R^{EV}_{b}\right) \leq \omega  \right]=\frac{1}{T-b+1}\sum_{t=1}^{T-b+1}I\left(\Gamma\left(R^{EV}_{b,t}\right)  \leq \omega\right).$$ 
Therefore, the $1-\alpha$th sample quantile, $c^{EV}_{T,b}(\alpha)$ defined as
$$
c^{EV}_{T,b}(\alpha) = \inf\left\{\omega: \hat{P}\left[\Gamma\left(R^{EV}_{b}\right) \leq \omega  \right] \geq 1-\alpha \right\},
$$
intuitively serves as the critical value for this test at the $\alpha$-level of the significance. The validity of this subsampling approach has been verified by \cite{escanciano2010specification}, who suggest an appropriate choice for bandwidth $b=\lfloor kT^{2/5}\rfloor$ for the sake of optimal minimax accuracy\footnote{$\lfloor z\rfloor$ denotes the largest integer that does not exceed z.}. Some numerical evidence has demonstrated that a diverse range of values of $k$, like 4, 5, and 6 provide reasonably good performance in finite samples. A centering strategy can be adopted for the resampling statistic to achieve better performance power-wise in finite samples.\newline  
Alternatively, \cite{escanciano2014specification} (hereafter EG) translate the restriction (\ref{specificationtest}) into 
\begin{equation}\label{eg}
    \E\left[\Psi_{\tau}\left(Y_t-\boldsymbol{X}'_t\boldsymbol{\theta}_0\right)\mathbb{I}\left(\boldsymbol{X}_t \leq x\right)\right]=0 \quad{} \forall x \in \mathbb{R}^{p+1},  \text{ for some } \boldsymbol{\theta}_0 \in \mathcal{B} \text{ and for all }\tau \in \Upsilon \subset (0,1).
\end{equation}
Naturally, a new test statistic can be constructed on the sample analog of moment conditions (\ref{eg}) with the replacement of $\boldsymbol{\theta}_0$ by their QAR estimates $\hat{\boldsymbol{\theta}}$,
\begin{equation}\label{egtest0}
T^{-1/2}\sum_{t=1}^{T}\Psi_{\tau}\left(Y_t-\boldsymbol{X}'_t\hat{\boldsymbol{\theta}}\right)  \mathbb{I}\left(\boldsymbol{X}_t \leq x\right) \quad{} \tau \in \Upsilon, x \in \mathbb{R}^{p+1}.
\end{equation}
Unlike the approach in the EV test, where the asymptotic behavior of the test statistic is derived by incorporating the non-negligible effect from the estimates of nuisance parameters into the final limiting distribution, \cite{escanciano2014specification} introduce a variant of weighting functions $\mathbb{I}\left(\boldsymbol{X}_t \leq x\right)$ satisfying the orthogonality condition for the Taylor expansion of the statistic (\ref{egtest0}) around the true parameter $\boldsymbol{\theta}_0$. With this consideration, the test statistic becomes   
\begin{equation}\label{egtest1}
    R^{EG}_{T}\left(x,\tau; \hat{\boldsymbol{\theta}}\right)=\sqrt{T}\sum_{t=1}^{T}\left\{\Psi_{\tau}\left(Y_t-\boldsymbol{X}'_t\hat{\boldsymbol{\theta}}\right)\left(\mathbb{I}\left(\boldsymbol{X}_{t}\leq x\right)-\hat{D}'_{T}\left(x,\hat{\boldsymbol{\theta}}(\tau)\right)\left(T^{-1}\sum_{t=1}^{T}\hat{\delta}_{t,\tau}\hat{\delta}'_{t,\tau}\right)^{-1}\hat{\delta}_{t,\tau}\right)\right\}
\end{equation}
with $\hat{D}_{T}\left(x,\hat{\boldsymbol{\theta}}(\tau)\right)= T^{-1}\sum_{t=1}^{T}\hat{\delta}_{t,\tau}\mathbb{I}\left(\boldsymbol{X}_{t}\leq x\right)$ and $$\hat{\delta}_{t,\tau}=\hat{f}\left(\boldsymbol{X}'_t\hat{\boldsymbol{\theta}}(\tau)\mid \boldsymbol{X}_t\right)\boldsymbol{X}_t,$$ where $\hat{f}\left(y\mid\boldsymbol{X}_t\right)$ is a consistent estimator of the conditional density function of $Y_t$ given the past information, $f\left(y\mid\boldsymbol{X}_t\right)$. One suggested kernel estimator is proposed by \cite{escanciano2012conditional}, defined by
\begin{equation}\label{kernel}
\hat{f}\left(\boldsymbol{X}'_t\hat{\boldsymbol{\theta}}(\tau)\mid \boldsymbol{X}_t\right) = \frac{1}{Mh_M}\sum_{j=1}^{M}K\left(\frac{\boldsymbol{X}'_t\hat{\boldsymbol{\theta}}(\tau)-\boldsymbol{X}'_t\hat{\boldsymbol{\theta}}(\tau_j)}{h_M}\right),
\end{equation}
where $\{{\tau_j}\}_{j=1}^{M}$ is a sequence randomly selected from $\Upsilon$ following the uniform distribution with $M \longrightarrow \infty$ as well as $T \longrightarrow \infty$. $K(\cdot)$ is a kernel function, and $h_M$ denotes a smoothing parameter which may depend on the data and the quantiles considered for the estimation. Compared to other density estimator candidates, this estimator is computationally less cumbersome but still preserves the same convergence rate as Rosenblatt estimator when the following assumption is imposed for the kernel function and the corresponding smoothing parameter.   
\begin{assumption}\label{ASS4}
\begin{enumerate}
    \item For kernel function K(s):
    \begin{enumerate}
        \item K(s) is continuously differentiable;
        \item $\int_{-\infty}^{\infty}K(s)=1$;
        \item K(s) is uniformly bounded;
        \item K(s) is of second order, i.e. $\int_{-\infty}^{\infty}sK(s)=0, \int_{-\infty}^{\infty}s^2K(s)ds \in (0,\infty)$ and $\int_{-\infty}^{\infty}K^2(s)ds \in (0,\infty)$.
    \end{enumerate}
    \item The convergence rate of smoothing parameter $h_M$ to 0 has to satisfy $P\left(a_M \leq h_M \leq b_M\right) \rightarrow 1$, for some deterministic sequences of positive numbers $a_M$ and $b_M$ such that $b_M \longrightarrow 0$ and $a^{p+2}_{M}M/logM \rightarrow \infty$ as $T \rightarrow \infty$.
\end{enumerate}
\end{assumption}
These regularity conditions for kernel functions apply to commonly used options in practice, for instance, the Gaussian kernel. 
\begin{corollary}[EG Test for non-causality]\label{egtest}
Under $\mathbb{H}_0$ and Assumptions \ref{ASS1}-\ref{ASS4}, let the matrix  $\E\left(\delta_{t,\tau}\delta'_{t,\tau}
\right)$ be nonsingular in a neighborhood of $\boldsymbol{\theta}(\tau)=\boldsymbol{\theta}_0(\tau)$ for all $\tau \in \Upsilon$,
\begin{equation*}
    R^{EG}_{T}\left(x,\tau; \hat{\boldsymbol{\theta}}\right) \Longrightarrow R^{EG}_{\infty}\left(x,\tau\right),
\end{equation*}
and 
\begin{equation*}
    \int_{\tau \in \Upsilon, x\in \mathbb{R}^{p+1}}\left\vert R^{EG}_T\left(x,\tau;\hat{\boldsymbol{\theta}}\right)\right\vert^2d\Phi(x)dW(\tau) \longrightarrow_{d} \int_{\tau \in \Upsilon, x\in \mathbb{R}^{p+1}}\left\vert R^{EG}_{\infty}\left(x,\tau\right)\right\vert^2d\Phi(x)dW(\tau),
\end{equation*}
where $R^{EG}_{\infty}$ is a Gaussian process with mean zero and covariance function characterized by 
$$
\left(\tau_1\wedge \tau_2-\tau_1\tau_2\right)\E\left\{\Pi_{\tau_1}\mathbb{I}\left(\boldsymbol{X}_t \leq x_1\right)\Pi_{\tau_2}\mathbb{I}\left(\boldsymbol{X}_t \leq x_2\right)\right\},
$$
with the so-called orthogonal projection operator on the weighting function $$\Pi_{\tau}\mathbb{I}\left(\boldsymbol{X}_t \leq x\right) \equiv \mathbb{I}\left(\boldsymbol{X}_t \leq x \right)-D'\left(x,\boldsymbol{\theta}_0(\tau)\right)\E^{-1}\left(\delta_{t,\tau}\delta_{t,\tau}\right)\delta_{t,\tau}.$$
and $\delta_{t,\tau}=f\left(\boldsymbol{X}'_t\boldsymbol{\theta}_0(\tau)\mid\boldsymbol{X}_t\right)\boldsymbol{X}_t$, $D\left(x,\boldsymbol{\theta}_0(\tau)\right)=\E\left(\delta_{t,\tau}\mathbb{I}\left(\boldsymbol{X}_t \leq x\right)\right)$. 
\end{corollary} 
As stated before, the main advantage of the EG test is the limiting distribution of the test statistic being invariant to the estimation effect of $\hat{\boldsymbol{\theta}}(\tau)$. Therefore, compared to the asymptotic distribution of the EV test, there is no "drift" term subtracted from a Gaussian process. However, this asymptotic distribution still depends on the data-generating process. Consequently, we cannot tabulate the critical values for the considered statistic. This is coped with the aid of a multiplier bootstrap approach. The approximation based on a transformation on $R^{EG}_{T}\left(x,\tau;\hat{\boldsymbol{\theta}}\right)$ is obtained by multiplying by a sequence of iid random variables $\{W_t\}_{t=1}^{T}$ with zero mean and unit variance, independent on $\boldsymbol{X}_t$,
\begin{equation}\label{transformed test}
\begin{split}
 & \tilde{R}^{EG}_{T,t}\left(x,\tau;\hat{\boldsymbol{\theta}}\right) \\
 = & \sqrt{T}\sum_{t=1}^{T}\left\{\Psi_{\tau}\left(Y_t-\boldsymbol{X}'_t\hat{\boldsymbol{\theta}}\right)\left(\mathbb{I}\left(\boldsymbol{X}_{t}\leq x\right)-\hat{D}'_{T}\left(x,\hat{\boldsymbol{\theta}}(\tau)\right)\left(T^{-1}\sum_{t=1}^{T}\hat{\delta}_{t,\tau}\hat{\delta}'_{t,\tau}\right)^{-1}\hat{\delta}_{t,\tau}\right)\right\}W_t.
 \end{split}
\end{equation}
One common choice of $\{W_t\}_{t=1}^{T}$ is 
\begin{equation}\label{multiplier}
\begin{cases}
 P\left(W=1-\omega\right)  = \omega/\sqrt{5} \quad{} \\
 P\left(W=\omega\right)  = 1-\omega/\sqrt{5},  \text{ with } \omega = \left(\sqrt{5}+1\right)/2.
\end{cases}
\end{equation}
This transformation has been proven by \cite{escanciano2014specification} to restore the limiting distribution of the original statistic. It allows us to use the empirical distribution of any continuous functional, including CvM form, $\Gamma \left(\tilde{R}^{EG}_{T,t}\right)$, i.e., $$\hat{P}\left[\Gamma\left(\tilde{R}^{EG}_{T,t}\right) \leq \omega \mid \{W_t\}_{t=1}^{T} \right] = \frac{1}{T}\sum_{t=1}^{T}I\left(\Gamma\left(\tilde{R}^{EG}_{T,t}\right) \leq \omega \right)$$
to consistently estimate the limiting distribution of the original statistic $\Gamma \left(R^{EG}_{T}\right)$. Likewise, the $(1-\alpha)$-th empirical quantile of the transformed statistic, $$c^{EG}_{T}(\alpha) = \inf\left\{\omega: \hat{P}\left[\Gamma \left(\tilde{R}^{EG}_{T}\right) \leq \omega \mid \{W_t\}_{t=1}^{T} \right] \geq 1-\alpha\right\},$$ will is a consistent estimate of the critical value at $\alpha$ significance level.\newline
In the presence of non-causality, $\psi(L^{-1})$ does not vanish from general MAR($r,s$) processes. Then, by Corollary \ref{col2},
\begin{equation*}\label{h1_ev}
    \E\left(\Psi_{\tau}\left(Y_t-\boldsymbol{X}'_t\boldsymbol{\theta}_1(\cdot)\right)\exp\left(i\cdot\boldsymbol{X}_t\right)\right) \neq 0 
\end{equation*}
and 
\begin{equation*}\label{h1_eg}
     \E\left(\Psi_{\tau}\left(Y_t-\boldsymbol{X}'_t\boldsymbol{\theta}_1(\cdot)\right)\mathbb{I}\left(\boldsymbol{X}_t \leq \cdot \right)\right) \neq 0
\end{equation*}
in a set with a positive Lebesgue measure on $\mathbb{R}^{p+1}\times \Upsilon$, provided that $\Phi$ and $W$ are absolutely continuous on $\mathbb{R}^{p+1}\times \Upsilon$ with respect to the Lebesgue measure. Correspondingly, under the alternative,
$$
\Gamma\left(R^{EV}_{T}\right)=\int_{\tau \in \Upsilon, x\in \mathbb{R}^{p+1}}\left\vert R^{EV}_T\left(x,\tau;\hat{\boldsymbol{\theta}}\right)\right\vert^2d\Phi(x)dW(\tau) \rightarrow_p \infty
$$
and 
$$
\Gamma\left(R^{EG}_{T}\right)=\int_{\tau \in \Upsilon, x\in \mathbb{R}^{p+1}}\left\vert R^{EG}_T\left(x,\tau;\hat{\boldsymbol{\theta}}\right)\right\vert^2d\Phi(x)dW(\tau) \rightarrow_p \infty,
$$
so both specification-based tests are consistent. 
\section{Monte Carlo Simulations}
\begin{figure}[htp]
    \centering
    \begin{subfigure}[h]{0.4\textwidth}
    \centering
    \includegraphics[width=\textwidth]{chi.png}
    \caption{$\chi^2_5-5$ distribution}
    \label{chi_qar}
    \end{subfigure}
\begin{subfigure}[h]{0.4\textwidth}
    \centering
    \includegraphics[width=\textwidth]{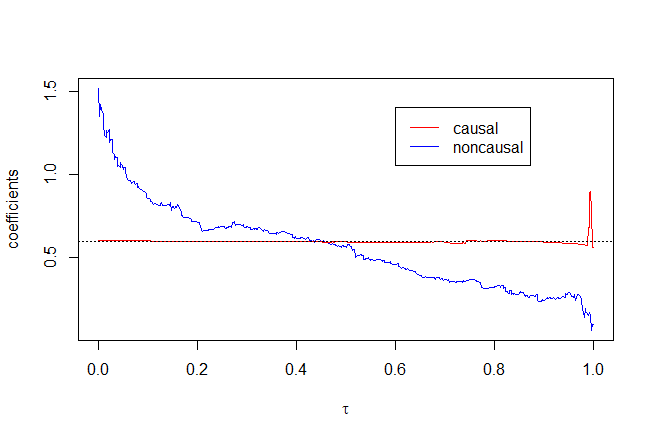}
    \caption{skewed normal distribution}
    \label{skewed_qar}
    \end{subfigure}
    \begin{subfigure}[h]{0.4\textwidth}
    \centering
    \includegraphics[width=\textwidth]{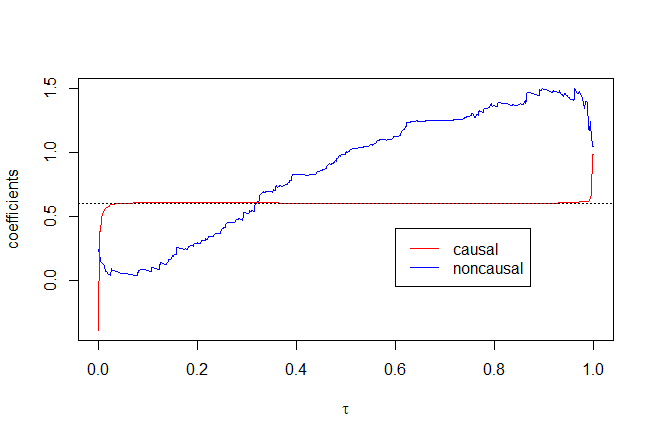}
    \caption{truncated Cauchy distribution}
    \label{truncated_qar}
    \end{subfigure}
    \begin{subfigure}[h]{0.4\textwidth}
    \centering
    \includegraphics[width=\textwidth]{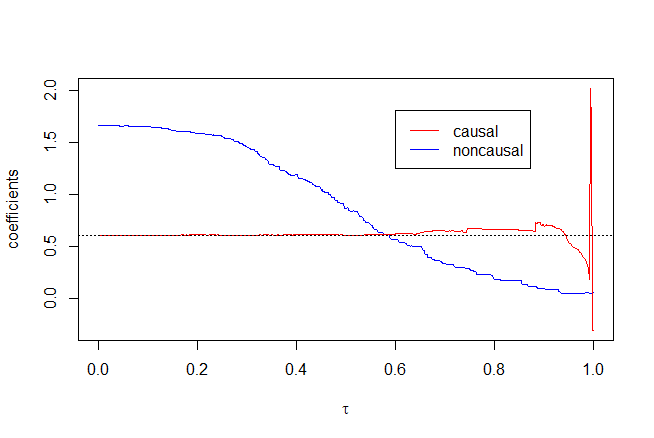}
    \caption{log normal distribution}
    \label{lognorm_qar}
    \end{subfigure}
    \caption{Four cases of QAR($1$) on MAR($1,0$) and MAR($0,1$), T=500, $\phi (\psi)=0.6$ : asymmetric distributions}
    \label{qar_plot_asymmetric}
\end{figure}

\begin{figure}[htp]
    \centering
    \begin{subfigure}[h]{0.4\textwidth}
    \centering
    \includegraphics[width=\textwidth]{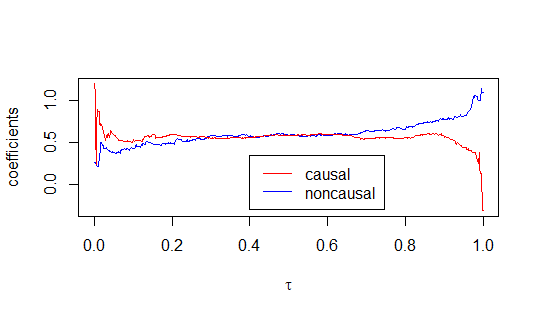}
    \caption{$t_3$ distribution}
    \label{t_qar}
    \end{subfigure}
\begin{subfigure}[h]{0.4\textwidth}
    \centering
    \includegraphics[width=\textwidth]{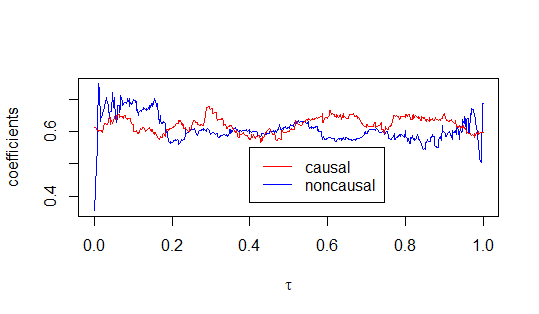}
    \caption{uniform distribution}
    \label{uniform_qar}
    \end{subfigure}
    \begin{subfigure}[h]{0.4\textwidth}
    \centering
    \includegraphics[width=\textwidth]{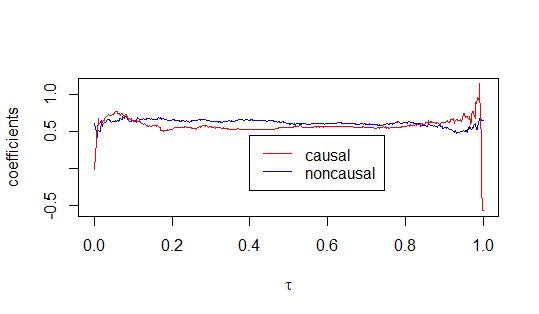}
    \caption{Laplace distribution}
    \label{laplace_qar}
    \end{subfigure}
    \caption{Three cases of QAR($1$) on MAR($1,0$) and MAR($0,1$), T=500, $\phi (\psi)=0.6$: symmetric distributions}
    \label{qar_plot_symmetric}
\end{figure}

In this section, we study the performance of the three proposed tests in finite samples and compare them with each other in terms of size and power.
\paragraph{Constancy Test}
In the first experiment, we focus on the approach based on the constancy test. In the simulation, we consider a pair of MAR($1,0$) and MAR($0,1$) models
\begin{equation}\label{simu1}
    \begin{cases}
    \left(1-\phi L\right)Y_t = u_t \\
    \left(1-\psi L^{-1}\right)Y^{*}_t = u_t, 
    \end{cases}
\end{equation}
which are generated from 11 non-Gaussian distributed innovations, which cover a majority of distributions commonly used in the empirics, ranging from symmetric to asymmetric, bounded to unbounded support, with mixed types of tail behavior. The parameters $(\phi,\psi)$ with values $(0.3, 0.6, 0.9)$ enable us to investigate the sensitivity of the method responding to data-generating processes with different persistence. The sample sizes are 200 and 500 with 500 replications. $\epsilon=0.05$ is the default choice for the quantile interval $[\epsilon,1-\epsilon] \subset (0,1)$. The empirical size and power of rejecting the constancy hypothesis to detect non-causality under the QAR framework are summarized in Table \ref{tab:constancy test}.\newline 
Regarding the size, the constancy test has an empirical size close to the nominal level in most cases but suffers from a severe over-rejection for heavy-tailed distributions. This corresponds to the estimation of conditional quantiles of these processes when $\tau$ is extremely close to $0$ or $1$, which generally calls for a larger sample size to produce less biased and more stable estimates. The volatility of QAR estimates of extremal quantiles triggers the over-rejection of the constant coefficients under the causality hypothesis. Therefore, as seen in Figures \ref{truncated_qar} and \ref{lognorm_qar}, where innovations follow truncated Cauchy distribution\footnote{Truncated at a sufficiently large value to ensure the existence of variance.} and log-normal distribution, respectively, the QAR estimates at quantiles close to 0 and 1 turn rather volatile compared to the estimates at other levels in $(0,1)$. To alleviate this issue, we propose to check different trimming strategies in the quantile interval for the test. There is an obvious trade-off in the selection of truncation of the quantile interval: over-trimmed, the power of the test will decrease due to the loss of valuable information; under-trimmed, the volatile estimate of extremal quantiles is not excluded, so the distortion in size will remain as before. Consequently, we conduct some experiments to analyze the sensitivity of the constancy test in response to truncated intervals; see Table \ref{tab:sensitivity of quantile intervel}. The results suggest a trimmed quantile interval $[0.10, 0.90]$ would be appropriate for the sample size considered because the power remains relatively high while the size is close to the nominal level. Another possible reason highlighted by \cite{koenker2002inference} is that the default smoothing parameter selection for estimating the density function of innovations, which comprises the test statistic in the \texttt{KhmaladzeTest} command in R studio, produces satisfactory performance for the class of distributions considered there, but is not designed for heavy-tailed distributed innovations. A more adaptive bandwidth choice of density function estimation of heavy-tailed distributions at extreme quantiles needs further investigation.\newline
From the perspective of power, the test achieves a significant leap in power as the sample size increases from 200 to 500 in most scenarios. More particularly, the method provides favorable performance in the presence of asymmetry in the distributions of innovations with rejection rates of $40\% \sim 90\%$ in 200-sized samples and $70\% \sim 90\%$ in 500-sized samples, respectively. This finding coincides with the idea shared in \cite{velasco2018frequency}, where the third-order moments contribute most to the identification of non-causal AR processes. This phenomenon is exceptionally well-illustrated in the first four cases (Exponential, Gamma, Beta, and F distributions) when the distribution of innovations is skewed but has no heavy tails. However, in the cases where innovations do not display skewness or heavy-tailedness, the test can barely distinguish non-causal processes from their causal counterparts, see Figure \ref{qar_plot_symmetric}. The point of the failure is illustrated in the same figure. The QAR estimates for non-causal AR($1$) with symmetric innovations appear to be indistinguishable from the ones for the causal counterparts, even under misspecification.  
\begin{table}[htp]
\centering
 \begin{adjustbox}{width=\textwidth}
 \begin{threeparttable}
\caption  {Empirical size and power of non-causality test using constancy$^{\star}$ test in QAR in various cases}
\label{tab:constancy test}
\begin{tabular}{@{} cccccccc @{}}
\toprule
  Distribution & test & \multicolumn{3}{c}{T=200}&  \multicolumn{3}{c}{T=500} \\ 
\hline
$u_t$ &    & $\phi (\psi)=0.3^{\dag}$   &  $\phi (\psi)=0.6$   & $\phi (\psi)=0.9$ & $\phi (\psi)=0.3$ & $\phi (\psi)=0.6$ & $\phi (\psi)=0.9$ \\    \midrule
 \multirow{2}{*}{Exp(1)-1}  & size &   4.20\% & 4.00\%  & 4.80\% & 6.40 \%  & 6.80\% & 4.80\%   \\
   &   power  & 33.80\% & 38.80\%  & 40.20\% & 65.40\% & 69.60\% & 72.40\% \\ \hline
    \multirow{2}{*}{Gamma(1,1)-1}  & size &   4.40\% & 3.00\%  & 3.80\% & 5.00 \%  & 5.80\% & 4.40\%   \\
   &   power  & 34.40\% & 37.00\%  & 42.20\% & 64.40\% & 68.40\% & 70.20\% \\ \hline
  \multirow{2}{*}{Beta(5,1)-5/6}  & size &   6.80\% & 6.60\%  & 5.40\% & 6.80 \%  & 9.40\% & 6.40\%   \\
   &   power  & 15.40\% & 27.20\%  & 39.00\% & 24.80\% & 65.00\% & 77.20\% \\ \hline
    \multirow{2}{*}{F(5,5)-5/3}  & size &   5.00\% & 6.60\%  & 3.40\% & 8.40 \%  & 6.00\% & 4.20\%   \\
   &   power  & 80.00\% & 81.80\%  & 70.00\% & 97.20\% & 98.20\% & 96.20\% \\ \hline
    \multirow{2}{*}{$\chi^2_5-5$}  & size &   5.00\% & 4.40\%  & 4.00\% & 4.00 \%  & 4.80\% & 5.20\%   \\
   &   power  & 11.40\% & 11.40\%  & 8.80\% & 27.40\% & 35.60\% & 17.40\% \\ \hline
    \multirow{2}{*}{skewed normal}  & size &   5.20\% & 6.60\%  & 7.40\% & 9.80 \%  & 7.40\% & 9.60\%   \\
   &   power  & 8.00\% & 7.80\%  & 10.60\% & 25.40\% & 20.00\% & 12.80\% \\ \hline
    \multirow{2}{*}{truncated Cauchy}  & size &   26.60\% & 34.80\%  & 43.00\% & 20.80 \%  & 19.20\% & 33.20\%   \\
   &   power  & 70.00\% & 96.40\%  & 92.40\% & 80.00\% & 99.80\% & 91.80\% \\ \hline
    \multirow{2}{*}{log normal}  & size &   21.20\% & 19.60\%  & 23.60\% & 23.20 \%  & 26.40\% & 26.60\%   \\
   &   power  & 98.20\% & 99.00\%  & 99.60\% & 100.00\% & 100.00\% & 100.00\% \\ \hline
    \multirow{2}{*}{$t_3$}  & size &   5.40\% & 4.00\%  & 5.00\% & 4.00 \%  & 6.80\% & 4.80\%   \\
   &   power  & 9.80\% & 13.00\%  & 15.80\% & 11.60\% & 20.00\% & 25.80\% \\ \hline
    \multirow{2}{*}{uniform}  & size &   6.20\% & 7.60\%  & 6.00\% & 6.00 \%  & 6.80\% & 5.80\%   \\
   &   power  & 13.40\% & 8.00\%  & 13.20\% & 40.40\% & 7.80\% & 48.80\% \\ \hline
    \multirow{2}{*}{Laplace}  & size &   3.40\% & 4.00\%  & 4.60\% & 2.60 \%  & 3.80\% & 4.60\%   \\
   &   power  & 8.40\% & 5.60\%  & 17.00\% & 10.80\% & 8.60\% & 29.00\% \\



\bottomrule
\end{tabular}
\begin{tablenotes}
\item $\star$: constancy test over the quantile interval $[0.05, 0.95]$. \\
\item $\dag$: $\phi(\psi)=0.3$ means the coefficient in the lag polynomial of the MAR($1,0$) process (purely causal) is $0.3$; the coefficient in the lead polynomial of the MAR($0,1$) process (purely non-causal) is $0.3$ as well, for comparison.    \\
\end{tablenotes}
\end{threeparttable}
\end{adjustbox}
\end{table}
\begin{table}[ht]
\centering
 \begin{adjustbox}{width=\textwidth}
 \begin{threeparttable}
\caption  {Empirical size and power of non-causality test using constancy test with different trimmed quantile interval}
\label{tab:sensitivity of quantile intervel}
\begin{tabular}{@{} cccccccc @{}}
\toprule
\hline
  Sample size & $[0.05, 0.95]$ & $\phi (\psi)=0.3$&  $\phi (\psi)=0.6$ & $\phi (\psi)=0.9$ & $\phi (\psi)=-0.4$& $\phi (\psi)=-0.6$&  $\phi (\psi)=-0.8$ \\  
\hline \hline
 \multirow{2}{*}{T=100} &  size  & 2.40\%  &  2.20\%   & 2.60\% & 2.80\% &  2.00\% & 2.40\% \\   
   &   power  & 22.40\% & 24.60\%  & 22.00\% & 4.80\% & 32.80\% & 26.60\% \\ \hline 
     \multirow{2}{*}{T=200} &  size  & 3.60\%  &  3.60\%   & 4.00\% & 5.00\% &  2.80\% & 4.40\% \\   
   &   power  & 40.80\% & 38.60\%  & 41.80\% & 8.40\% & 48.00\% & 44.00\% \\ \hline
  \multirow{2}{*}{T=500}  & size &   3.40\% & 6.20\%  & 4.80\% & 6.60 \%  & 4.80\% & 3.40\%   \\
   &   power  & 67.20\% & 67.00\%  & 72.40\% & 14.20\% & 66.00\% & 72.20\% \\ \hline\hline
   Sample size & $[0.10, 0.90] $ & $\phi (\psi)=0.3$&  $\phi (\psi)=0.6$ & $\phi (\psi)=0.9$ & $\phi (\psi)=-0.4$& $\phi (\psi)=-0.6$&  $\phi (\psi)=-0.8$ \\ \hline \hline 
   \multirow{2}{*}{T=100} &  size  & 3.40\%  &  3.20\%   & 3.20\% & 3.80\% &  2.40\% & 3.40\% \\   
   &   power  & 28.00\% & 25.40\%  & 22.80\% & 6.00\% & 32.80\% & 28.20\% \\ \hline
     \multirow{2}{*}{T=200} &  size  & 3.20\%  &  2.80\%   & 5.00\% & 3.00\% &  4.20\% & 4.00\% \\   
   &   power  & 37.40\% & 43.40\%  & 43.20\% & 5.80\% & 49.40\% & 48.80\% \\ \hline
  \multirow{2}{*}{T=500}  & size &   4.80\% & 5.40\%  & 6.00\% & 4.60 \%  & 3.80\% & 5.00\%   \\
   &   power  & 71.20\% & 67.80\%  & 69.20\% & 14.60\% & 65.40\% & 72.80\% \\ \hline\hline
    Sample size & $[0.15, 0.85] $ & $\phi (\psi)=0.3$&  $\phi (\psi)=0.6$ & $\phi (\psi)=0.9$ & $\phi (\psi)=-0.4$& $\phi (\psi)=-0.6$&  $\phi (\psi)=-0.8$ \\ \hline \hline 
   \multirow{2}{*}{T=100} &  size  & 4.40\%  &  2.60\%   & 3.60\% & 5.00\% &  3.00\% & 3.00\% \\   
   &   power  & 27.80\% & 28.20\%  & 21.80\% & 8.80\% & 37.60\% & 30.20\% \\ \hline
     \multirow{2}{*}{T=200} &  size  & 5.60\%  &  3.80\%   & 3.60\% & 6.60\% &  3.40\% & 4.40\% \\   
   &   power  & 39.00\% & 40.00\%  & 42.40\% & 9.60\% & 51.60\% & 46.60\% \\ \hline
  \multirow{2}{*}{T=500}  & size &   5.40\% & 4.40\%  & 5.00\% & 4.20 \%  & 4.80\% & 5.40\%   \\
   &   power  & 64.80\% & 65.80\%  & 69.00\% & 21.80\% & 68.00\% & 71.60\% \\ \hline \bottomrule
   



\end{tabular}
\begin{tablenotes}
\item The innovations follow exponential distribution.  \\
\end{tablenotes}
\end{threeparttable}
\end{adjustbox}
\end{table}
 %
\paragraph{EV Test} With the same setting as the one in the Constancy Test, we take the case with the coefficient $\phi (\psi)= 0.6$ for MAR($1,0$) (MAR($0,1$)) as the representative to examine the performance of EV test in discriminating non-causal processes from their causal alternatives. The sample size varies from 100 to 200 and 500. The number of replications is 500. Regarding the bandwidth for the subsampling scheme to approximate the critical value, we choose $b=\lfloor 4T^{2/5}\rfloor$. As for the weighting functions involved in the expression (\ref{ev_cvm}), a uniform distribution is applied to $W(\tau)$ defined on the evenly discretized quantile interval $\Upsilon = [0.01, 0.99]$ and 2-dimensional standard normal distribution to $\Psi(x)$ for the sake of simplicity in the calculation. The results of the empirical size and power are displayed in Table \ref{tab:ev test}. Concerning size, the results present some fluctuation around the nominal level. The test tends to under-reject the correct null hypothesis in the light-tailed scenarios while over-rejects in the heavy-tailed scenarios. This comes from the subjective choice in the subsampling size, whose performance can be sensitive to the quantile interval included in the test and the data-generating process. Nevertheless, the distortion in size is not so significant, and we believe this can be eased by choosing different subsampling schemes. As for power, the test achieves reasonably good performance generally. From the simulation results, we observe that the power of detecting non-causality is close to $100\%$ in most cases when the sample size is 500, which is as expected. It is worth noting that the convergence rate of power towards $100\%$ varies across different distributions. The numerical evidence indicates that the further the innovations depart from Gaussianity, the faster the convergence rate is. In such log-normal and uniform distributions, a reasonably high power, like $86\%$ or $92\%$, has been obtained in relatively small samples. While others (chi-square) may need larger samples to reach the same level of power.\newline
\begin{table}[ht]
\centering
 \begin{adjustbox}{width=0.6\textwidth}
 \begin{threeparttable}
\caption  {Empirical size and power of non-causality test using EV test in QAR in various cases}
\label{tab:ev test}
\begin{tabular}{@{} lcccccccccccc @{}}
\toprule
  Distribution & \multicolumn{3}{r}{parameter:} & \multicolumn{9}{c}{$\phi (\psi)=0.6$}   \\ 
\cline{5-12} 
       $u_t$ &    &    &   \multicolumn{1}{r}{T:}   & \multicolumn{2}{c}{100} &  & \multicolumn{2}{c}{200}  &  & \multicolumn{2}{c}{500} & \\    \midrule
 \multirow{2}{*}{Gaussian}      &    & size &  & \multicolumn{2}{c}{2.20\%} &  & \multicolumn{2}{c}{3.40\%} &  & \multicolumn{2}{c}{3.40\%} &   \\
   &  & power &  & \multicolumn{2}{c}{0.40\%} &  & \multicolumn{2}{c}{0.60\%} &  & \multicolumn{2}{c}{2.60\%} &         \\  \hline        
   \multirow{2}{*}{exponential} &  & size  &  & \multicolumn{2}{c}{4.40\%} & & \multicolumn{2}{c}{1.80\%} &  & \multicolumn{2}{c}{2.00\%} &  \\
  &  & power &  & \multicolumn{2}{c}{29.20\%} &  & \multicolumn{2}{c}{43.20\%} & &   \multicolumn{2}{c}{97.20\%} &    \\ \hline
   \multirow{2}{*}{Gamma} &  & size  &  & \multicolumn{2}{c}{1.80\%} & & \multicolumn{2}{c}{3.00\%} &  & \multicolumn{2}{c}{2.80\%} &  \\
  &  & power &  & \multicolumn{2}{c}{10.60\%} &  & \multicolumn{2}{c}{41.60\%} & &   \multicolumn{2}{c}{97.60\%} &    \\ \hline
   \multirow{2}{*}{Beta} &  & size  &  & \multicolumn{2}{c}{3.20\%} & & \multicolumn{2}{c}{3.20\%} &  & \multicolumn{2}{c}{3.40\%} &  \\
  &  & power &  & \multicolumn{2}{c}{26.40\%} &  & \multicolumn{2}{c}{67.80\%} & &   \multicolumn{2}{c}{99.00\%} &    \\ \hline
   \multirow{2}{*}{F} &  & size  &  & \multicolumn{2}{c}{3.60\%} & & \multicolumn{2}{c}{5.80\%} &  & \multicolumn{2}{c}{3.40\%} &  \\
  &  & power &  & \multicolumn{2}{c}{24.00\%} & & \multicolumn{2}{c}{67.60\%} & &   \multicolumn{2}{c}{99.00\%} &    \\ \hline
    \multirow{2}{*}{$\chi^2_5-5$} &  & size  &  & \multicolumn{2}{c}{4.80\%} & & \multicolumn{2}{c}{4.40\%} &  & \multicolumn{2}{c}{3.60\%} &  \\
  &  & power &  & \multicolumn{2}{c}{13.40\%} &  & \multicolumn{2}{c}{22.20\%} & &   \multicolumn{2}{c}{68.40\%} &    \\ \hline
   \multirow{2}{*}{log normal} &  & size  &  & \multicolumn{2}{c}{5.40\%} & & \multicolumn{2}{c}{4.60\%} &  & \multicolumn{2}{c}{6.80\%} &  \\
  &  & power &  & \multicolumn{2}{c}{54.00\%} &  & \multicolumn{2}{c}{92.40\%} & &   \multicolumn{2}{c}{100.00\%} &    \\ \hline
    \multirow{2}{*}{$t_3$} &  & size  &  & \multicolumn{2}{c}{6.20\%} & & \multicolumn{2}{c}{7.00\%} &  & \multicolumn{2}{c}{6.40\%} &  \\
  &  & power &  & \multicolumn{2}{c}{13.80\%} &  & \multicolumn{2}{c}{42.40\%} & &   \multicolumn{2}{c}{93.40\%} &    \\ \hline
    \multirow{2}{*}{Uniform} &  & size  &  & \multicolumn{2}{c}{5.00\%} & & \multicolumn{2}{c}{6.00\%} &  & \multicolumn{2}{c}{3.40\%} &  \\
  &  & power &  & \multicolumn{2}{c}{44.00\%} &  & \multicolumn{2}{c}{86.80\%} & &   \multicolumn{2}{c}{100.00\%} &    \\ \hline
   \multirow{2}{*}{Laplace} &  & size  &  & \multicolumn{2}{c}{4.40\%} & & \multicolumn{2}{c}{4.80\%} &  & \multicolumn{2}{c}{4.40\%} &  \\
  &  & power &  & \multicolumn{2}{c}{14.00\%} &  & \multicolumn{2}{c}{47.20\%} & &   \multicolumn{2}{c}{98.20\%} &    \\ \hline


\bottomrule
\end{tabular}
\begin{tablenotes}
\item EV test: the choice of size for subsampling is $\lfloor4T^{2/5} \rfloor$.  \\
\end{tablenotes}
\end{threeparttable}
\end{adjustbox}
\end{table}

\paragraph{EG Test}
The setup of DGP keeps unchanged, like in the EV test. The sample size varies from 50 to 100 and 200. The weighting functions $\Psi(x)$ and $W(\tau)$ in the CvM form of $R^{EG}_{T}\left(x,\tau;\hat{\boldsymbol{\theta}}\right)$ are chosen to be the empirical distribution of $\boldsymbol{X}_t$ and uniform distribution over the grid of quantiles from $\Upsilon=[0.01,0.99]$ considered in the estimation. The critical value is obtained through the multiplier bootstrap introduced in the methodology section. Implementing the multiplier bootstrap avoids computing the estimates for each subsample. The results are summarized in Table \ref{tab:eg test}. Regarding the empirical size performance, this approach delivers stable rejection frequencies under the null hypothesis associated with their nominal level in all cases considered in the simulation exercise. This is anticipated in accordance with the argument in the previous section that there is no subjective choice involved in the approximation of the critical value. In terms of power, the EG test has an increasing trend as the sample is enlarged. Similar to the EV test, the EG approach outperforms in the presence of skewness and excess kurtosis (regardless of negative or positive), with a rejection probability over 70\% in relatively small samples (200). For the cases where the performance is less satisfactory, such as $t_3$ and Laplace distributions, power still increases when the sample size becomes larger.\newline
\begin{table}[ht]
\centering
 \begin{adjustbox}{width=0.6\textwidth}
 \begin{threeparttable}
\caption  {Empirical size and power of non-causality test using EG test in QAR in various cases}
\label{tab:eg test}
\begin{tabular}{@{} lcccccccccccc @{}}
\toprule
  Distribution & \multicolumn{3}{r}{parameter:} & \multicolumn{9}{c}{$\phi (\psi)=0.6$}   \\ 
\cline{5-12} 
       $u_t$ &    &    &   \multicolumn{1}{r}{T:}   & \multicolumn{2}{c}{50} &  & \multicolumn{2}{c}{100}  &  & \multicolumn{2}{c}{200} & \\    \midrule
 \multirow{2}{*}{Gaussian}      &    & size &  & \multicolumn{2}{c}{6.00\%} &  & \multicolumn{2}{c}{5.00\%} &  & \multicolumn{2}{c}{4.00\%} &   \\
   &  & power &  & \multicolumn{2}{c}{4.80\%} &  & \multicolumn{2}{c}{5.60\%} &  & \multicolumn{2}{c}{5.80\%} &         \\  \hline        
   \multirow{2}{*}{exponential} &  & size  &  & \multicolumn{2}{c}{5.20\%} & & \multicolumn{2}{c}{4.60\%} &  & \multicolumn{2}{c}{6.00\%} &  \\
  &  & power &  & \multicolumn{2}{c}{24.80\%} &  & \multicolumn{2}{c}{49.20\%} & &   \multicolumn{2}{c}{76.40\%} &    \\ \hline
   \multirow{2}{*}{Gamma} &  & size  &  & \multicolumn{2}{c}{5.00\%} & & \multicolumn{2}{c}{5.60\%} &  & \multicolumn{2}{c}{5.00\%} &  \\
  &  & power &  & \multicolumn{2}{c}{23.80\%} &  & \multicolumn{2}{c}{45.80\%} & &   \multicolumn{2}{c}{75.80\%} &    \\ \hline
   \multirow{2}{*}{Beta} &  & size  &  & \multicolumn{2}{c}{6.40\%} & & \multicolumn{2}{c}{5.80\%} &  & \multicolumn{2}{c}{6.00\%} &  \\
  & & power & & \multicolumn{2}{c}{24.20\%} &  & \multicolumn{2}{c}{42.20\%} & &   \multicolumn{2}{c}{75.80\%} &    \\ \hline
   \multirow{2}{*}{F} &  & size  &  & \multicolumn{2}{c}{4.40\%} & & \multicolumn{2}{c}{5.00\%} &  & \multicolumn{2}{c}{5.40\%} &  \\
  &  & power &  & \multicolumn{2}{c}{22.60\%} & & \multicolumn{2}{c}{37.40\%} & &   \multicolumn{2}{c}{67.00\%} &    \\ \hline
    \multirow{2}{*}{$\chi^2_5-5$} &  & size  &  & \multicolumn{2}{c}{6.60\%} & & \multicolumn{2}{c}{5.40\%} &  & \multicolumn{2}{c}{4.60\%} &  \\
  &  & power &  & \multicolumn{2}{c}{12.80\%} &  & \multicolumn{2}{c}{22.60\%} & &   \multicolumn{2}{c}{41.60\%} &    \\ \hline
   \multirow{2}{*}{log normal} &  & size  &  & \multicolumn{2}{c}{5.60\%} & & \multicolumn{2}{c}{5.40\%} &  & \multicolumn{2}{c}{4.00\%} &  \\
  &  & power &  & \multicolumn{2}{c}{32.00\%} &  & \multicolumn{2}{c}{60.60\%} & &   \multicolumn{2}{c}{85.80\%} &    \\ \hline
    \multirow{2}{*}{$t_3$} &  & size  &  & \multicolumn{2}{c}{4.40\%} & & \multicolumn{2}{c}{6.60\%} &  & \multicolumn{2}{c}{7.00\%} &  \\
  &  & power &  & \multicolumn{2}{c}{6.80\%} &  & \multicolumn{2}{c}{14.80\%} & &   \multicolumn{2}{c}{36.20\%} &    \\ \hline
    \multirow{2}{*}{Uniform} &  & size  &  & \multicolumn{2}{c}{4.80\%} & & \multicolumn{2}{c}{6.60\%} &  & \multicolumn{2}{c}{6.60\%} &  \\
  &  & power &  & \multicolumn{2}{c}{10.20\%} &  & \multicolumn{2}{c}{25.40\%} & &   \multicolumn{2}{c}{64.80\%} &    \\ \hline
   \multirow{2}{*}{Laplace} &  & size  &  & \multicolumn{2}{c}{5.60\%} & & \multicolumn{2}{c}{5.00\%} &  & \multicolumn{2}{c}{6.80\%} &  \\
  &  & power &  & \multicolumn{2}{c}{8.20\%} &  & \multicolumn{2}{c}{14.40\%} & &   \multicolumn{2}{c}{41.60\%} &    \\ \hline


\bottomrule
\end{tabular}
\begin{tablenotes}
\item EG test: critical value obtained from multiplier bootstrap. \\
\end{tablenotes}
\end{threeparttable}
\end{adjustbox}
\end{table}
An overall comparison of the three methods is exhibited in Table \ref{tab:comparson}. Size-wise, the EG test has an appealing attribute of undistorted size in general scenarios compared to the other two approaches. On the contrary, the constancy test suffers from over-rejection in heavy-tailed cases, and the EV test delivers less accuracy than the EG test in some cases. Power-wise, the EG test is the most robust one as it produces relatively good results in all situations but is extraordinarily competent for asymmetric distributions. By contrast, the EV test achieves the highest power among the three in the symmetric cases. In comparison, the constancy test can be a powerful tool in detecting non-causality in processes with heavy tails. However, given that the consistency of the constancy test for non-causality is not guaranteed and the size is distorted, it may give misleading conclusions in empirical analysis. Thus, the constancy test can only be considered a preliminary test to check whether the process is likely non-causal, followed by the implementation of the EV or EG tests, which serve as the formal tests for non-causality. In practice, a combination of the constancy test and EV (EG) test is suggested. \newline
\begin{table}[ht]
\centering
 \begin{adjustbox}{width=\textwidth}
 \begin{threeparttable}
\caption  {Comparison of QAR-based non-causality tests}
\label{tab:comparson}
\begin{tabular}{@{} lcccccccccccc @{}}
\toprule
  Distribution & \multicolumn{3}{r}{test type:} & \multicolumn{2}{c}{constancy test}&  &\multicolumn{2}{c}{EV test} &  & \multicolumn{2}{c}{EG test}   \\ 
\cline{5-6} \cline{8-9} \cline{11-12}
       $u_t$ &    &    &   \multicolumn{1}{r}{T:}   & 100 & 200 &  & 100 & 200  &  & 100 & 200 & \\    \midrule
 \multirow{2}{*}{Gaussian}      &    & size &  & 2.80\% & 5.40\% &  & 2.20\% & 3.40 \% &  & 5.00\% & 4.00\% &   \\
   &  & power &  & 4.20\% & 3.40\% &  & 0.40\% & 0.60\% &  & 5.60\% & 5.80\% &         \\  \hline        
   \multirow{2}{*}{exponential} &  & size  &  & 3.20\% & 4.00\% & & 4.40\% & 1.80\% &  & 4.60\% & 6.00\% &  \\
  &  & power &  & 25.40\% & 38.80\% &  & 29.20\% & 43.20\% & &   49.20\% & 76.40\% &    \\ \hline
\multirow{2}{*}{Gamma} &  &  size & & 3.80\% & 3.00\% &  & 1.80\% & 3.00\% & & 5.60\% & 5.00\% &   \\
 & & power &  & 26.60\% & 37.00\% & & 10.60\% & 41.60\% & & 45.80\% & 75.80\% & \\
 \hline
\multirow{2}{*}{Beta} &  &  size & & 4.20\% & 6.60\% &  & 3.20\% & 3.20\% & & 5.80\% & 6.00\% &   \\
 & & power &  & 18.60\% & 27.20\% & & 26.40\% & 67.80\% & & 42.20\% & 75.80\% & \\
 \hline
\multirow{2}{*}{F} &  &  size & & 6.80\% & 6.60\% &  & 3.60\% & 5.80\% & & 5.00\% & 5.40\% &   \\
 & & power &  & 62.40\% & 81.80\% & & 24.00\% & 67.60\% & & 34.70\% & 67.00\% & \\
 \hline
 \multirow{2}{*}{$\chi^2_5$} &  &  size & & 5.80\% & 4.40\% &  & 4.80\% & 4.40\% & & 5.40\% & 4.60\% &   \\
 & & power &  & 9.60\% & 11.40\% & & 13.40\% & 22.20\% & & 22.60\% & 41.60\% & \\
 \hline
 \multirow{2}{*}{log normal} &  &  size & & 20.20\% & 19.60\% &  & 5.40\% & 4.60\% & & 5.40\% & 4.00\% &   \\
 & & power &  & 78.80\% & 99.60\% & & 54.00\% & 92.40\% & & 60.60\% & 85.80\% & \\
 \hline
 \multirow{2}{*}{$t_3$} &  &  size & & 3.40\% & 4.00\% &  & 6.20\% & 7.00\% & & 6.60\% & 7.00\% &   \\
 & & power &  & 10.20\% & 13.00\% & & 13.80\% & 42.40\% & & 14.80\% & 36.20\% & \\
 \hline
 \multirow{2}{*}{Uniform} &  &  size & & 6.80\% & 7.60\% &  & 5.00\% & 6.00\% & & 6.60\% & 6.60\% &   \\
 & & power &  & 7.40\% & 8.00\% & & 44.00\% & 86.80\% & & 25.40\% & 64.80\% & \\
 \hline
 \multirow{2}{*}{Laplace} &  &  size & & 5.80\% & 4.00\% &  & 4.40\% & 4.80\% & & 5.00\% & 6.80\% &   \\
 & & power &  & 5.20\% & 5.60\% & & 14.00\% & 47.20\% & & 14.40\% & 41.60\% & \\


\bottomrule
\end{tabular}
\begin{tablenotes}
\item Constancy test: with trimmed quantile interval $[0.05,0.95]$
\end{tablenotes}
\end{threeparttable}
\end{adjustbox}
\end{table}
\section{Empirical Applications}
\begin{figure}[htp]
    \centering
    \includegraphics[width=\textwidth]{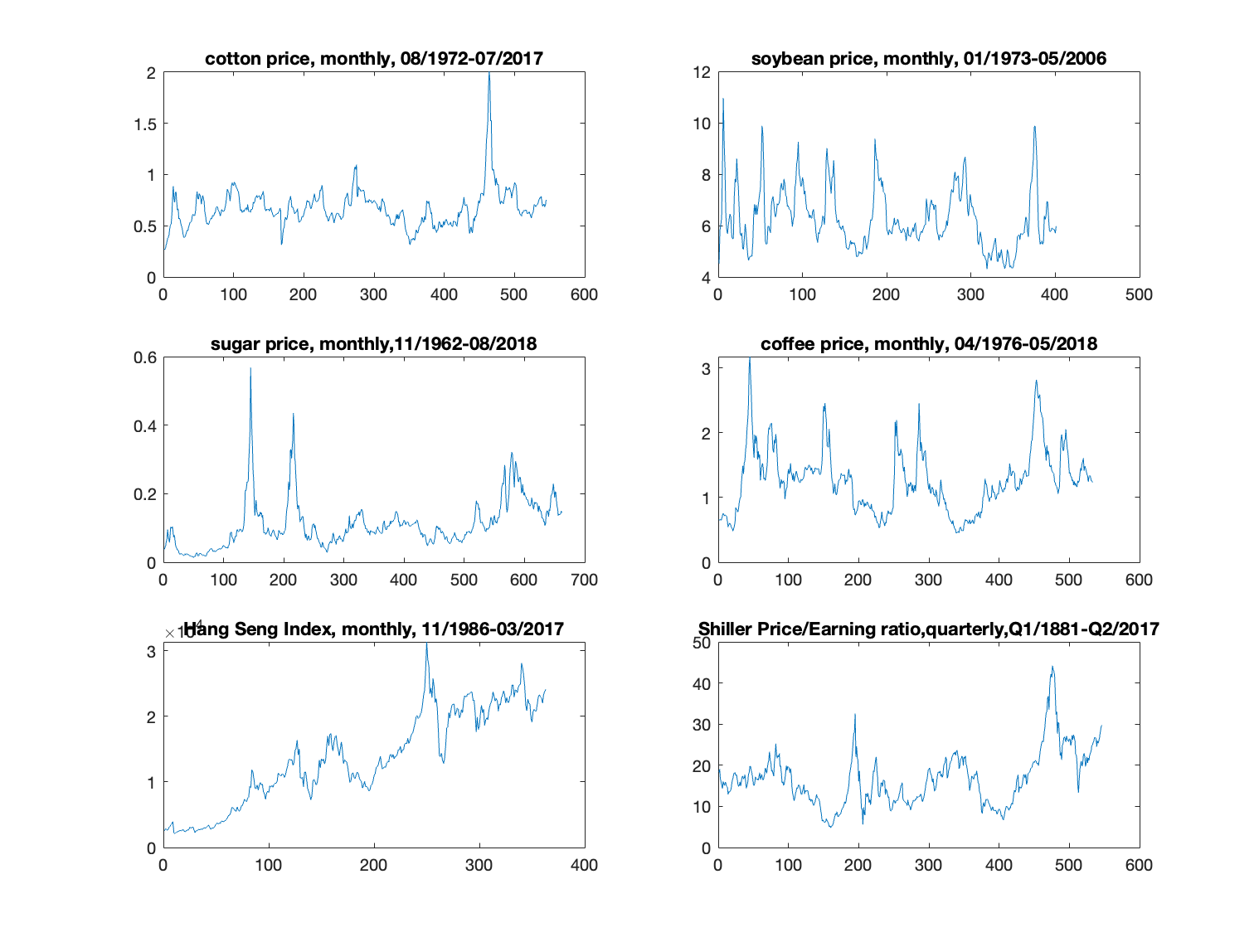}
    \caption{Financial series trajectories}
    \label{financial series}
\end{figure}

\begin{figure}[htp]
    \centering
    \includegraphics[width=\textwidth]{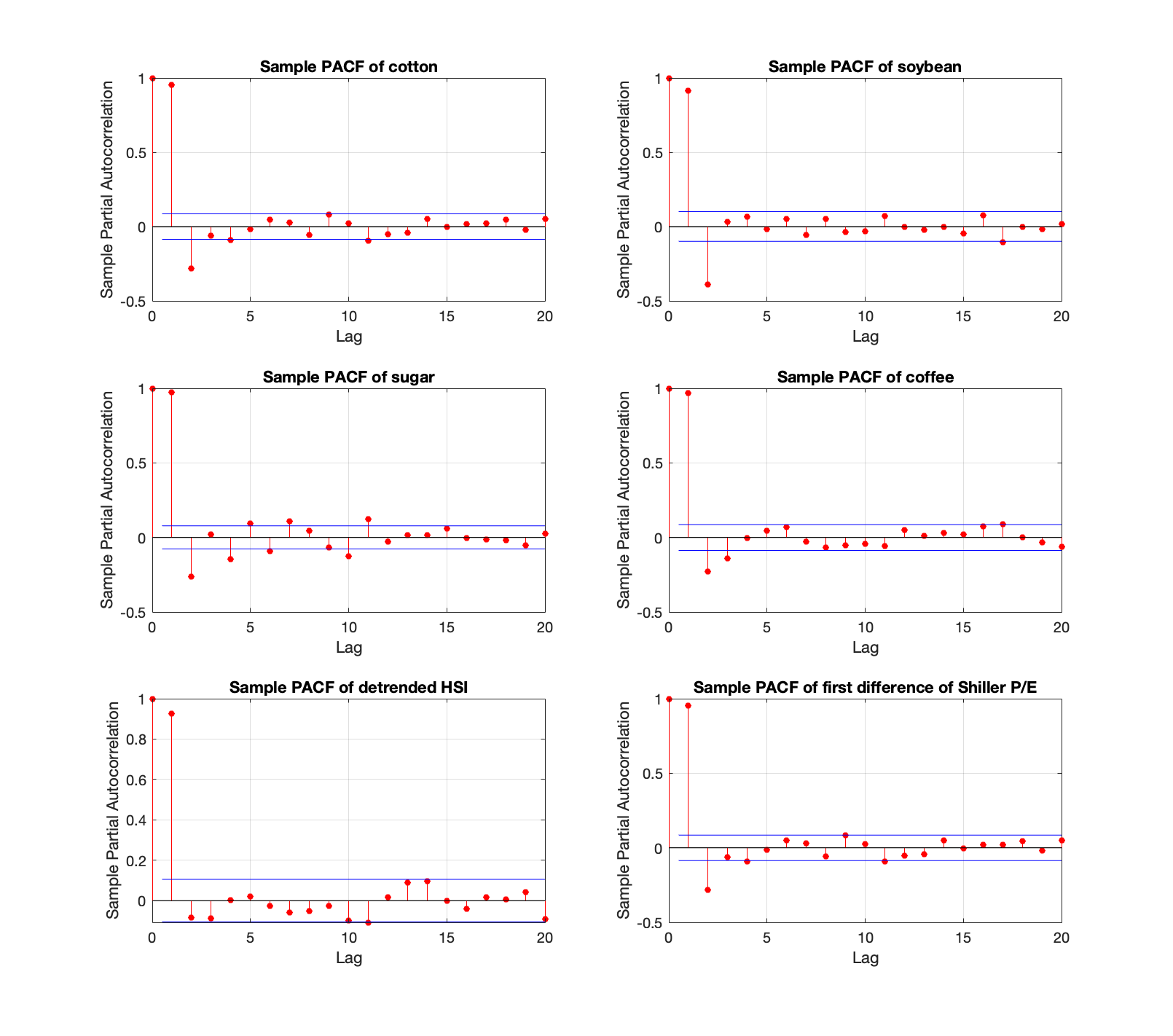}
    \caption{Sample partial autocorrelation functions of six financial series}
    \label{financial series pacf}
\end{figure}

In this section, we apply our non-causality tests to six financial series studied in \cite{fries2019mixed}: cotton price, soybean price, sugar price, coffee price, Hang Seng Index (HSI), and Shiller Price/Earning ratio (Shiller PE), where single or multiple spikes and asymmetric dynamics are exhibited. \cite{fries2019mixed} found numerical evidence in favor of MAR($r,s$) models in fitting time series with local explosiveness phases compared to purely causal AR models. The frequency of data is quarterly for the Shiller PE series and monthly for the rest\footnote{The access of the replication data can be found in \cite{fries2019mixed}}. The trajectories of these series are displayed in Figure \ref{financial series}.\newline
Before proceeding to our testing strategies, we must ensure that the series is stationary. The augmented Dickey-Fuller tests indicate no unit roots in the cotton, soybean, sugar, and coffee price. Neither the evidence of unit root is found in the series of HSI after a linear trend is subtracted. The stationarity of the Shiller PE series is secured after the first difference in the levels. The sample partial autocorrelation function for each series is computed, see Figure \ref{financial series pacf}, to determine the order of the lag orders: cotton: 2; soybean: 2; sugar: 4; coffee: 3; HSI: 1; Shiller PE: 7.\newline
Three non-causality testing procedures: the constancy test, the EV test, and the EG test, are applied to each series, respectively. The corresponding results are reported in Table \ref{tab:financial series test}. For the constancy test, we consider two trimmed quantile intervals: $[0.05, 0.95]$ and $[0.10, 0.90]$ to mitigate the instability effect on the power from the subjective choice in the quantiles. Both in the cases of cotton and sugar series, a significant fluctuation in the coefficients is observed. Yet no strong evidence against linear quantile specification is found based on the EV or EG test. The strong rejection in the constancy test for these series may result from the over-rejection issue in heavy-tailed scenarios, which makes the results from the EV or EG tests more reliable. This does not deviate much from the results obtained by \cite{fries2019mixed}, where they found one non-causal root (0.94) in the cotton series and  a root (0.92) in the sugar series. As seen in the numerical experiments, a non-causal AR model with coefficients near the unity makes it harder to distinguish from its causal counterpart, as well as generates less distinct dynamics than its causal counterpart. For three series (soybean, coffee, and Shiller PE), the tests show strong evidence favoring non-causal processes at 5\% or even 1\% level from all three testing strategies. For HSI, the test shows mild evidence of non-causality at the significance level of 10\% based on the EG test. This result is compatible with the conclusion drawn in \cite{fries2019mixed}, where mixed models with nontrivial non-causal components\footnote{The coefficients of the non-causal components in the MAR($r,s$) models are not closed to the unity.} are selected. 

\begin{table}[ht]
\centering
 \begin{adjustbox}{width=\textwidth}
 \begin{threeparttable}
\caption  {Non-causality tests for six financial series in \cite{fries2019mixed} }
\label{tab:financial series test}
\begin{tabular}{@{} lcccccccccccc @{}}
\toprule
  Financial series & \multicolumn{3}{r}{test type:} & \multicolumn{2}{c}{constancy test}&  &\multicolumn{2}{c}{EV test} &  & \multicolumn{2}{c}{EG test}   \\ 
\cline{5-6} \cline{8-9} \cline{11-12}
        &  \multicolumn{1}{c}{total AR order$(r+s)$}   &    &     & $[0.05,0.95]^{\dag}$ & $[0.10,0.90]$ &  &  &   &  & &  & \\    \midrule
\multirow{2}{*}{Cotton}      & \multirow{2}{*}{2}    & statistic &  & $6.897^{**}$ & $4.738^{**}$ &  & \multicolumn{2}{c}{0.012} &  & \multicolumn{2}{c}{0.025} &   \\
   &  & critical value\footnote{the default level of significance is 5\%} & & (3.393) & (3.287) &  & \multicolumn{2}{c}{(0.046)} &  & \multicolumn{2}{c}{(0.032)} &         \\  \hline        
   \multirow{2}{*}{Soybean} &  \multirow{2}{*}{2} & statistic  &  & $4.703^{**}$ & $4.445^{**}$ & & \multicolumn{2}{c}{$0.209^{**}$} &  & \multicolumn{2}{c}{$0.027^{**}$} &  \\
  &  & critical value &  & (3.393) & (3.287) &  & \multicolumn{2}{c}{(0.158)} & &   \multicolumn{2}{c}{(0.021)}&    \\ \hline
\multirow{2}{*}{Sugar} & \multirow{2}{*}{4} &  statistic & & $306.043^{**}$ & $144.025^{**}$ &  & \multicolumn{2}{c}{0.009} & & \multicolumn{2}{c}{0.065} &   \\
 & & critical value &  & (5.560) & (5.430) & & \multicolumn{2}{c}{(0.102)} & & \multicolumn{2}{c}{(0.100)} & \\
 \hline
\multirow{2}{*}{coffee} & \multirow{2}{*}{3} &  statistic & & $8.457^{**}$ & $6.992^{**}$ &  & \multicolumn{2}{c}{$0.149^{**}$} & & \multicolumn{2}{c}{$0.044^{**}$} &   \\
 & & critical value &  & (4.523) & (4.383) & & \multicolumn{2}{c}{(0.089)} & & \multicolumn{2}{c}{(0.025)}& \\
 \hline
\multirow{2}{*}{Hang Seng Index} & \multirow{2}{*}{1} &  statistic & & 0.017 & 0.012 &  & \multicolumn{2}{c}{0.168} & & \multicolumn{2}{c}{$0.078^{*}$} &   \\
 & & critical value &  & (2.140) & (2.102) & & \multicolumn{2}{c}{(0.176)} & & \multicolumn{2}{c}{(0.083)} & \\
 \hline
 \multirow{2}{*}{Shiller's P/E ratio} & \multirow{2}{*}{7}  &  statistic & & $15.105^{**}$ & $9.027^{**}$ &  & \multicolumn{2}{c}{$0.197^{**}$} & & \multicolumn{2}{c}{$0.135^{**}$} &   \\
 & & critical value &  & (8.578) & (8.368) & & \multicolumn{2}{c}{(0.189)} & & \multicolumn{2}{c}{(0.066)}&  \\


\bottomrule
\end{tabular}
\begin{tablenotes}
\item $\dag$ the trimmed quantile interval considered in the constancy test. \\
\item $**$ stands for significance at level $5\%$ and $*$ stands for significance at $10\%$. \\
\item the critical value at $10\%$ significance level for the EG test in the case of Hang Seng Index is 0.061. 

\end{tablenotes}
\end{threeparttable}
\end{adjustbox}
\end{table}
\section{Extensions and conclusion}
\subsection{Some Extensions}
So far, the preceding discussion has been confined to MAR($r,s$) driven by $iid$ innovations. Within this framework, the only possible source of non-linearity in MAR($r,s$) is non-causality, which contributes to the consistency of the test in the aforementioned methods. However, stylized nonlinear dynamics like conditional heteroskedasticity or asymmetric dynamics are prevalently observed in the financial and macroeconomic data, which renders it more demanding to detect non-causality in a time series process. A more robust methodology applicable to AR models accommodating non-linear features needs investigation. The critical point is how to disentangle non-linearity induced by non-causality from the other alternatives. If these non-linear features can be captured by a parametric model, one plausible solution is to incorporate these non-linear terms into the baseline model (\ref{mixed autoregression}). Below we list some possible extensions where this strategy is employed.
\paragraph{Asymmetric Dynamics}
This can be solved by allowing varying coefficients in the MAR($r,s$) model in the spirit of the random coefficient model, defined by
\begin{equation}\label{random MAR}
    \tilde{\phi}(L)\tilde{\psi}(L^{-1})Y_t = u_t
\end{equation}
where $\tilde{\phi}(L)=1-\phi_1(U_t)L-\dots-\phi_{r}(U_t)L^{r}$ and $\tilde{\psi}= 1-\psi_{1}(U_t)L^{-1}-\dots-\psi_{s}(U_t)L^{-s}$, $U_t$ is an $iid$ sequence of random variables following standard uniform distribution, and $u_t$ is an $iid$ innovation sequence satisfying Assumptions \ref{ASS1}. Denote $$
\Omega_{c} = \begin{pmatrix}
\phi_1(U_t) & \dots & \phi_{r-1}(U_t) & \phi_{r}(U_t) \\
& \boldsymbol{I}_{r-1}& & \boldsymbol{0}_{(r-1)\time 1}
\end{pmatrix}
$$
and 
 $$
\Omega_{nc} = \begin{pmatrix}
\psi_1(U_t) & \dots & \psi_{s-1}(U_t) & \psi_{s}(U_t) \\
& \boldsymbol{I}_{s-1}& & \boldsymbol{0}_{(s-1)\time 1}
\end{pmatrix}.
$$
Similar to the $p$-th order autoregressive process, which is designed to accommodate asymmetric dynamics in \cite{koenker2006quantile} for linear QAR model, we need to assume $\E\left(\Omega_{c} \otimes \Omega_{c} \right)$ and $\E\left(\Omega_{nc} \otimes \Omega_{nc} \right)$ have eigenvalues with moduli less than one. This equation (\ref{random MAR}) is able to mimic asymmetric dynamics since $\phi_j, \psi_j$'s are functions $[0,1] \rightarrow \mathbb{R}$. The definition of non-causality in this context will be adapted to that $\tilde{\psi}(L^{-1})$ does not decline to constant. In the causal situation, this model (\ref{random MAR}) works like a random coefficient model with lags. The linearity of the conditional mean is restored, and the linear quantile dynamic model with varying coefficients over different quantiles remains the correct specification for conditional quantiles of $Y_t$. On the other hand, when the process is non-causal, it is conceivable that linearity will not hold anymore. Therefore, the methodology relying on the specification tests is applicable here.\newline
\paragraph{Volatility Clustering}
Concerning volatility clustering, which is routinely modeled by the quadratic ARCH/GARCH model in squared residuals. Another popular choice is to replace the squared value with the absolute value suggested by \cite{taylor2008modelling} and make the model a linear ARCH.
\begin{equation}\label{linear arch}
    \begin{split}
        & \phi(L)\psi(L^{-1})Y_t = v_t \\
        \text{ where } & v_t = \sigma_t u_t \\
        & \sigma_t = \gamma_0+ \gamma_1 |v_{t-1}|+\dots+ \gamma_q |v_{t-q}|
    \end{split}
\end{equation}
where $\phi(L)$ and $\psi(L^{-1})$ are defined following (\ref{mixed autoregression}), and $u_t$ is a sequence of iid innovations. The linear ARCH model is able to capture the correlation in the variance and meanwhile preserves a relatively simple linear specification compared to other alternatives like GARCH. Under the $\mathbb{H}_0$ where $\psi(L^{-1})$ degenerates to 1, the linearity of conditional quantile specification still holds after adding $\{|v_{t-j}|\}_{j=1}^{q}$ into the regression equation.
\begin{equation}\label{linear arch qar}
\begin{split}
   Q_{Y_t}\left(\tau|Y_{t-1},Y_{t-2},\dots \right) & = \phi_{1}Y_{t-1}+\dots+\phi_{r}Y_{t-r}+ Q_{u_t}(\tau)\left(\gamma_0+\gamma_1|v_{t-1}|+\dots+ \gamma_q |v_{t-q}|\right) \\
   & = \underbrace{Q_{u_t}(\tau)\gamma_0}_{\tilde{\gamma}_0(\tau)}+\underbrace{Q_{u_t}(\tau)\gamma_1}_{\tilde{\gamma}_1(\tau)}|v_{t-1}|+\dots+\underbrace{Q_{u_t}(\tau)\gamma_q}_{\tilde{\gamma}_q(\tau)}|v_{t-q}|+\phi_{1}Y_{t-1}+\dots+\phi_{r}Y_{t-r} \\
   & = \tilde{\gamma}_0(\tau)+\tilde{\gamma}_1(\tau)|v_{t-1}|+\dots+\tilde{\gamma}_q(\tau)|v_{t-q}|+\phi_1Y_{t-1}+\dots+\phi_{r}Y_{t-r},
    \end{split}
\end{equation}
where $|v_{t-j}|$ can be recovered by $Y_{t-j}-\phi_{1}Y_{t-j-1}-\dots-\phi_{r}Y_{t-j-r}$. Under non-causality, the explicit expression of the conditional quantile of $Y_t$ remains unclear. Nevertheless, it cannot be characterized by encompassing linear combinations of residuals in the model. Consequently, the linear dynamic quantile model would not be the correct specification conceivably.\newline 
Overall, these two possible extensions to cases with nonlinear dynamics are tentative since the statistical properties of conditional quantiles of $Y_t$ defined by (\ref{random MAR}) and (\ref{linear arch}) require further investigation, which opens a couple of lines for future research. Some simulation trials in Appendix \ref{extensionsimulation} have shown the validity of the proposed strategies.
\subsection{Perspective from Extreme Quantiles}
One intriguing observation from the simulation is that QAR estimates at extreme quantiles might be informative for identifying the true models even though linear quantile specification is incorrect for the conditional quantile of non-causal processes. As depicted in Figure \ref{qar_plot_asymmetric}, for MAR($0,1$) processes with coefficient $0.6$ driven from asymmetric innovations,   
\begin{equation*}
    \begin{split}
       \left(1-0.6 L^{-1}\right)Y_t = u_t  & \Leftrightarrow  Y_t = (0.6)^{-1}Y_{t-1}-(0.6)^{-1}u_{t-1}. \\
    \end{split}
\end{equation*}
The estimated slope of $Y_{t-1}$ approaches $(0.6)^{-1}$ when the quantile gets close to 0 or 1. Somehow it indicates that the linear correlation at extreme quantiles between $Y_t$ and 
$Y_{t-1}$ can help to discriminate causality and non-causality. A similar idea has been adopted for model selection based on the extreme clustering of residuals by \cite{fries2019mixed}, but the method is restricted to $\alpha-$stable distributions. Rich data is required for further analysis to get a less biased estimator for conditional quantiles close to 0 or 1. This opens a possible avenue for future research in line with identifying causal and non-causal processes using tail information of processes.  
\subsection{Conclusion}
This paper introduces three novel testing strategies for non-causality in linear time series within the quantile regression framework. The tests exploit the non-linearity of autoregressive processes with non-causality and achieve the objective of detecting non-causality based on the well-developed inference under the QAR framework. The constancy test shares the simplicity of implementation but lacks consistency since the behavior of linear quantile autoregression for non-causal processes is not clear yet. This issue from the constancy test is tackled by testing the specification of linear conditional quantile models to detect non-causality. Specification-based non-causality testing procedures like EV and EG tests yield stable size at a nominal level and fairly satisfactory power. On the one hand, the EV test outperforms the EG test in platykurtic and leptokurtic situations or symmetric distributions. On the other hand, the EG test is less computationally cumbersome and shows better performance when the process is skewed. However, no method is placed in a dominating situation. Thus, a combined testing procedure with the constancy test as a preliminary check complemented with either EV or EG test is suggested for practitioners.\newline
Some possible extensions to accommodate different dependence in model innovations, which might bring obstacles in detecting non-causality, are proposed at the end of the paper. Some simulation results in QAR estimate at extreme quantiles indicate the possibility of identifying non-causal processes by employing information from the tails of processes.  
\clearpage


\clearpage

\phantomsection
\addcontentsline{toc}{section}{References}  
\makeatletter
\makeatother

\bibliographystyle{rss}
\bibliography{quantile}
\newpage
\section{Appendices}
\subsection{\textbf{Some Properties of Non-causal Autoregressive Processes}}
\paragraph{Higher-order Dependence of All-pass Time Series Models}\label{higher order dependence}
Following the same setup in Example \ref{ex3}, 
\begin{equation*}
    \tilde{u}_t = \frac{1-\psi L}{1-\psi L^{-1}}u_t =\sum_{j=-\infty}^{\infty}\rho_ju_{t+j}.
\end{equation*}
The skewness of $\tilde{u}_t$ is 
\[
\E\left(\tilde{u}^3_t\right)=\sum_{j=-\infty}^{\infty}\rho^3_j\E\left(u^3_t\right)=\left(1-\frac{3\psi^2(\psi+1)}{\psi^2+\psi+1}\right) \E\left(u^3_t\right), |\psi|<1,
\]
where $-1<\left(1-\frac{3\psi^2(\psi+1)}{\psi^2+\psi+1}\right)<1$.
From the above expression, it is easy to tell that the all-pass filter preserves the symmetry of the innovations if the original ones are not skewed but might alter the direction of skewness if $u_t$ is asymmetric by varying values of $\psi$. Apart from the correlation in the squared value of $\tilde{u}_t$ that has been explicitly shown in Example \ref{ex3}, here we derive the closed-form solution for the dependence at order 3. It suffices to show $\Cov\left(\tilde{u}^3_t, \tilde{u}^3_{t+h}\right)$ is nonzero for $h \neq 0$.
\begin{equation*}
    \begin{split}
        &\E\left(\tilde{u}^3_t\tilde{u}^3_{t+h}\right)  = \E\left(\sum_{j=-\infty}^{\infty}\sum_{i=-\infty}^{\infty}\sum_{m=-\infty}^{\infty}\sum_{n=-\infty}^{\infty}\sum_{l=-\infty}^{\infty}\sum_{k=-\infty}^{\infty}\rho_j\rho_i\rho_m\rho_n\rho_l\rho_ku_{t+j}u_{t+i}u_{t+m}u_{t+h+n}u_{t+h+k}u_{t+h+l}\right) \\
        =& \left(\sum_{j=-\infty}^{\infty}\rho^3_j\rho^3_{j+h}\right)\left(\E(u^6_t)-15\E(u^4_t)\E(u^2_t)-10\E^2(u^3_t)-15\E^3(u^2_t)\right)\\
        & + 3\left(\sum_{j=-\infty}^{\infty}\left(\rho_{j+h}\rho^{3}_j+\rho^3_{j+h}\rho_j\right)\right)\E(u^4_t)\E(u^2_t) \\
        & + \left(\left(\sum_{j=-\infty}^{\infty}\rho^3_j\right)^2+9\left(\sum_{j=-\infty}^{\infty}\rho^2_{j+h}\rho_{j}\right)\left(\sum_{j=-\infty}^{\infty}\rho_{j+h}\rho^2_{j}\right)\right)\E^2(u^3_t)
    \end{split}
\end{equation*}
after using $\sum_{j=-\infty}^{\infty}\rho_j\rho_{j+h}=0$ (coincides with no correlation property of all-pass time series process) and $\sum_{j=-\infty}^{\infty}\rho^2_j=1$ (variance preserving property),
\begin{equation*}
     \Cov\left(\tilde{u}^3_t, \tilde{u}^3_{t+h}\right)  = \E\left(\tilde{u}^3_t\tilde{u}^3_{t+h}\right)-\E\left(\tilde{u}^3_t\right)\E\left(\tilde{u}^3_{t+h}\right)
\end{equation*}
generally is not zero. For instance, $h=1$, after simplification
\[
\Cov\left(\tilde{u}^3_t, \tilde{u}^3_{t+1}\right)=\alpha_1\E(u^6_t)+\left(\alpha_2\E(u^4_t)+\alpha_4\E^2(u^2_t)\right)\E(u^2_t)+\alpha_3\E^2(u^3_t)
\]
with
\begin{equation*}
    \begin{cases}
    \alpha_1 = \frac{-3\psi^5(1-\psi^2)^3}{\psi^4+\psi^2+1} \\
    \alpha_2 = -3\psi^3(1-\psi^2)^3+\frac{45\psi^5(1-\psi^2)^3}{\psi^4+\psi^2+1} \\
    \alpha_3 =\frac{30\psi^5(1-\psi^2)^3}{\psi^4+\psi^2+1} - \frac{9(1-\psi^2)^3(2\psi+1)\psi^2}{(\psi^2+\psi+1)^2} \\
    \alpha_4 = \frac{45\psi^5(1-\psi^2)^3}{\psi^4+\psi^2+1}
    \end{cases}
\end{equation*}
where zeros are not attained at the same value of $\psi \in (-1,1)\setminus\{0\}$. 
\paragraph{Conditional Density Function of Non-causal Autoregressions}\label{conditional density function}
In this section, we try to study the properties of the conditional density function of the response variable in the presence of non-causality in the autoregressive process through simulations. Consider a pair of MAR($1,0$) and MAR($0,1$) processes generated from $iid$ innovations following the same distribution with the density function $f_u(\cdot)$,
\begin{equation}\label{cdsimulation}
    \begin{cases}
    Y_t = 0.6Y_{t-1}+u_t \\
    \tilde{Y}_t = 0.6^{-1}\tilde{Y}_{t-1}+u_t = 0.6\tilde{Y}_{t+1}-0.6u_{t+1}.
    \end{cases}
\end{equation}
One is purely causal, and the other is purely non-causal with a coefficient of $0.6$. We start by analyzing $f\left(Y_{t}\leq y|Y_{t-1}=x\right)$ in the causal case. 
\begin{equation}\label{conditionaldc}
    \begin{split}
      f\left(Y_t = y|Y_{t-1}=x\right) & =\frac{d P\left(Y_t \leq y |Y_{t-1}=x\right)}{dy} \\ 
      & = \frac{d P\left(0.6Y_{t-1}+u_t \leq y |Y_{t-1}=x\right)}{dy} \\
      & = \frac{d P\left(u_t \leq y-0.6x |Y_{t-1}=x\right)}{dy} \\
      & = f_u\left(y-0.6x\right)=f_u\left(y-0.6Y_{t-1}\right), 
    \end{split}
\end{equation}
It is not difficult to conclude from equation \ref{conditionaldc} that the conditional density of $Y_t$ given $Y_{t-1}$ is shifting horizontally as the value of $Y_{t-1}$ varies. Despite the change in the location of the density function, the rest remains the same across different values of $Y_{t-1}$.\newline
Similarly, we derive the conditional density function for the non-causal case,
\begin{equation}\label{cdensitync}
    \begin{split}
        f\left(\tilde{Y}_t=y|\tilde{Y}_{t-1}=x\right) & = \frac{f\left(\tilde{Y}_{t-1}=x|\tilde{Y}_t=y\right)f\left(\tilde{Y}_t=y\right)}{f(\tilde{Y}_{t-1}=x)} \text{ by Bayes rule }\\
        & = \frac{f\left(0.6\tilde{Y}_{t}-0.6u_t=x\mid \tilde{Y}_{t}=y\right)f\left(\tilde{Y}_t=y\right)}{f\left(Y_{t-1}=x\right)} \text{ by definition of $\tilde{Y}_{t-1}$} \\
        & = \frac{f_u\left(y-0.6^{-1}x\right)f\left(\tilde{Y}_t=y\right)}{f\left(Y_{t-1}=x\right)}  \text{ by the independence of $u_t$ and $\tilde{Y}_t$} \\
        & =\frac{f_u\left(y-0.6^{-1}x\right)f\left(\tilde{Y}_t=y\right)}{\int_{-\infty}^{\infty}f\left(\tilde{Y}_{t-1}=x|\tilde{Y}_t=s\right)f\left(\tilde{Y}_t=s\right)ds} \text{ law of total probability }\\
        &=\frac{f_u\left(y-0.6^{-1}x\right)f\left(\tilde{Y}_t=y\right)}{\int_{-\infty}^{\infty}f_u\left(s-0.6^{-1}x\right)f\left(\tilde{Y}_t=s\right)ds}.
    \end{split}
\end{equation}
There is no general closed-form solution to this expression. Note that no $x$ plays a role in $f(Y_t=y)$, and the denominator is $x-$dependent but highly nonlinear due to the integration. If we assign additive property\footnote{Additive property states the sum of independent variables from the same distribution would follow the distribution from the same family. The common distributions which share this property are $\alpha$-stable distribution, exponential distribution, geometric distribution, etc.} to the marginal distribution of $Y_t$. This nonlinearity can be shown more clearly. Say $u_t$ follows an exponential distribution with rate $\lambda$, then the equation \ref{cdensitync} has the explicit form
\begin{equation*}
    \begin{split}
        & f\left(Y_t=y|Y_{t-1}=x\right)\\
        = & \frac{\left(\int_{-\infty}^{\infty}e^{-isy}\prod_{j=0}^{\infty}\frac{\lambda}{\lambda-is(0.6)^j}ds\right)\lambda e^{-(x-0.6y)}}{\left(\int_{-\infty}^{\infty}e^{-isx}\prod_{j=0}^{\infty}\frac{\lambda}{\lambda-is(0.6)^j}ds\right)}\mathbb{I}\left(x-0.6y \geq 0\right).
    \end{split}
\end{equation*}
This suggests that the shape (functional form) of the density function would differ, corresponding to the choice of $x$. The following simulation experiment demonstrates this argument.
\begin{figure}[h]
    \centering
    \begin{subfigure}[h]{0.4\textwidth}
    \centering
    \includegraphics[width=\textwidth]{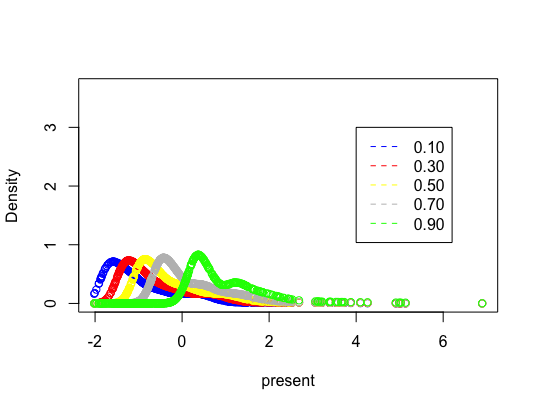}
    \caption{conditional density of $Y_t$ in the causal case}
    \label{cdc}
    \end{subfigure}
\begin{subfigure}[h]{0.4\textwidth}
    \centering
    \includegraphics[width=\textwidth]{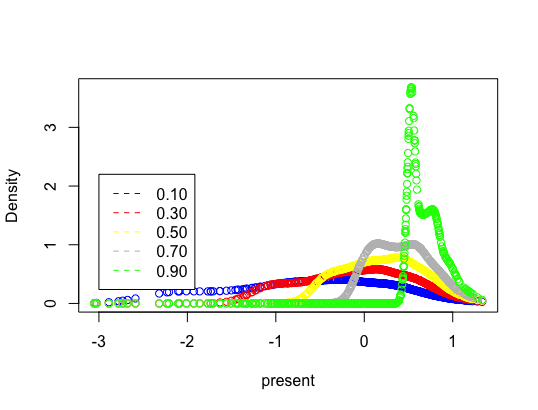}
    \caption{conditional density of $\tilde{Y}_t$ in the non-causal case}
    \label{cdnc}
    \end{subfigure}
    \caption{conditional density of $Y_t$ given different $x$}
    \label{cd}
\end{figure}
In this simulation, we generate two AR(1) processes (\ref{cdsimulation}) by exponentially distributed innovations with rate 1. The sample size is 500. Estimated conditional density functions $f\left(y|Y_{t-1}=x\right)$ are plotted in Figure \ref{cd}, given five choices of $x$: $10\%$, $30\%$, $50\%$, $70\%$, and $90\%$ percentiles of $Y_{t-1}$ sample. The density function is estimated by $akj$ command in R Studio, which is a univariate adaptive kernel estimation used by \cite{portnoy1989adaptive}. The left panel displays the result for the causal case. As shown in (\ref{conditionaldc}), density functions in different colors (values of $x$) share the same shape but the location. Whereas in the right panel, where the estimated density is plotted, five estimated conditional density functions present distinct modes, skewness, and kurtosis.
\subsection{Asymptotic Properties of QAR Estimates}\label{qarestimate}
QAR is first proposed by \cite{koenker2006quantile} to study the conditional quantile functions of the following $p$th-order AR process,
\begin{equation}\label{qar}
    Y_t= \theta_0(U_t)+\theta_1(U_{t})Y_{t-1}+\cdots+\theta_p(U_t)Y_{t-p},
\end{equation}
where $U_t$ is an iid sequence distributed as a standard uniform. This expression can be regarded as an AR($p$) process allowing coefficients of lags to be random but somewhat dependent on each other. By the property of monotone transformation, our target, the conditional quantile at each $\tau \in (0,1)$, can be written as
\begin{equation}\label{conditionalquantile}
    Q_{Y_t}\left(\tau|Y_{t-1},\dots,Y_{t-p}\right)= \theta_0(\tau)+\theta_1(\tau)Y_{t-1}+\dots+\theta_p(\tau)Y_{t-p}, \quad{} \tau \in (0,1).
\end{equation}
The estimates of $\boldsymbol{\theta}(\tau)=\left(\theta_0(\tau),\theta_1(\tau),\dots,\theta_p(\tau)\right)'$ are obtained by minimizing the following objective function,
\begin{equation}\label{lossfunctionqar}
\hat{\boldsymbol{\theta}}(\tau) = \argmin_{\boldsymbol{\theta}\in \mathbb{R}^{p+1}}\sum_{t=1}^{T}\rho_{\tau}(Y_t-\boldsymbol{X}'_t\boldsymbol{\theta}),
\end{equation}
where $\boldsymbol{X}'_t=(1,Y_{t-1},\dots,Y_{t-p})$ and check function $\rho_{\tau}(u)=u\left(\tau-\mathbb{I}(u<0)\right)$. A vectorized form of (\ref{qar}) is introduced to facilitate the asymptotic analysis of the estimates $\hat{\boldsymbol{\theta}}(\tau)$,
\begin{equation*}
    \boldsymbol{Y}_t = \boldsymbol{A}_t\boldsymbol{Y}_{t-1}+\boldsymbol{V}_t,
\end{equation*}
with
\begin{equation*}
    \boldsymbol{A}_t = \begin{pmatrix}
    \theta_1(U_t) & \theta_2(U_t) & \dots & \theta_{p}(U_t) \\
    & \boldsymbol{I}_{p-1}& & \boldsymbol{0}_{(p-1)\times1}
    \end{pmatrix} \text{ and } \boldsymbol{V}_t = \begin{pmatrix}
    \epsilon_t \\
    \boldsymbol{0}_{(p-1)\times1}
    \end{pmatrix}
\end{equation*}
where $\epsilon_t=\theta_0(U_t)-\E\left(\theta_0(U_t)\right)$ and $\boldsymbol{Y}_t=\left(Y_t, Y_{t-1},\dots, Y_{t-p+1}\right)'$. The study of asymptotic properties is based on the following conditions
\begin{enumerate}
    \item $\{\epsilon_t\}$ are iid innovations with mean 0 and finite variance $\sigma^2<\infty$. The distribution function of $\epsilon_t$, F, admits a continuous density $f(\epsilon)$ away from zero on $E=\{\epsilon: 0<F(\epsilon)<1\}$.
    \item The eigenvalues of $\E\left(\boldsymbol{A}_t\otimes\boldsymbol{A}_t\right)$ have moduli within unity.
    \item The conditional distribution function $P(Y_t<\cdot|Y_{t-1},Y_{t-2},\dots)$ denoted by $F_{t-1}(\cdot)$ has a density function $f_{t-1}(\cdot)$ uniformly integrable on $E$. 
\end{enumerate}
Under these three assumptions, 
\begin{equation*}\label{qarasymptotics}
\Sigma^{-1/2}\sqrt{T}\left(\hat{\boldsymbol{\theta}}(\tau)-\boldsymbol{\theta}(\tau)\right) \rightarrow_d \boldsymbol{B}_{p+1}(\tau)
\end{equation*}
where $\boldsymbol{B}_{k}(\tau)$ is a $k$-dimensional Brownian bridge. By definition it can be written as $\mathcal{N}(\boldsymbol{0}, \tau(1-\tau)\boldsymbol{I}_k)$ for any given $\tau$. $\Sigma$ is a matrix characterized by density and distribution function of $\epsilon_t$. Let $\Sigma_0= \E\left(\boldsymbol{X}_t \boldsymbol{X}'_t\right)$ and $\Sigma_1= \E\left(f_{t-1}\left(F^{-1}_{t-1}(\tau)\right)\boldsymbol{X}_t \boldsymbol{X}'_t\right)$. Then $\Sigma$ is defined as $\Sigma^{-1}_1\Sigma_0\Sigma^{-1}_1$.\newline
In the special case where the data generating process is a conventional causal AR model with fixed coefficients, the conditional density would be independent of $\boldsymbol{X}_t$. We will have $\Sigma_{1}=f(F^{-1}(\tau))\E\left(\boldsymbol{X}_t \boldsymbol{X}'_t\right)$. Further we can simplify the $\Sigma$ to $f^{-2}(F^{-1}(\tau))\E^{-1}\left(\boldsymbol{X}_t \boldsymbol{X}'_t\right)$.  
\subsection{Proof to Theorem \ref{linearity}}  
\paragraph{Non-causality $\Rightarrow$ Nonlinear conditional quantile}
We prove this statement by contradiction. Consider a stationary MAR($r,s$) with innovations satisfying Assumption \ref{ASS1} and $s>0$. Assume all conditional quantiles of $Y_t$ are linear in $\left(Y_{t-1}, Y_{t-2},\dots,Y_{t-p}\right)$, where $p=r+s$. That is, 
$$
Q_{Y_t}\left(\tau|Y_{t-1}, Y_{t-2},\dots,Y_{t-p}\right)= \theta_0(\tau)+\theta_1(\tau)Y_{t-1}+\dots+\theta_p(\tau)Y_{t-p} \text{ for any given }\tau \in (0,1).
$$
By aggregating $Q_{Y_t}\left(\tau|Y_{t-1}, Y_{t-2},\dots, Y_{t-p}\right)$ over the entire quantile range, the linearity is maintained for the aggregation. Therefore, we have 
\begin{equation}\label{contradiction}
\int_{0}^{1}Q_{Y_t}\left(\tau|Y_{t-1}, Y_{t-2},\dots,Y_{t-p}\right)d\tau = \int_{0}^{1}\theta_0(\tau)d\tau + \int_{0}^{1}\theta_1(\tau)d\tau Y_{t-1}+\dots+\int_{0}^{1}\theta_{p}(\tau)d\tau Y_{t-p}    
\end{equation}
Equivalently, we can yield
\begin{equation*}
    \E\left(Y_t|Y_{t-1}, Y_{t-2},\dots,Y_{t-p}\right)= \theta_0+\theta_1Y_{t-1}+\dots+\theta_p Y_{t-p},
\end{equation*}
which contradicts the statement in Corollary 5.2.3 by \cite{rosenblatt2000gaussian} on the nonlinearity of conditional expectation in past information with the presence of non-causality. Thus, the presumed statement is not valid. That is to say, there exists a conditional quantile of $Y_t$ which is nonlinear in the past information for at least one $\tau \in (0,1)$ if $s>0$. 
\paragraph{Nonlinear conditional quantile $\Rightarrow$ Non-causality} 
This can be demonstrated equivalently by its contrapositive statement: an AR process $Y_t$ being causal implies its conditional quantile $Q_{Y_t}\left(\tau\mid I_{t-1}\right)$ is linear for all $\tau \in (0,1).$ \newline
If a MAR($r,s$) is purely causal, i.e., s=0 and r=p. Then we have 
\begin{equation*}
    Y_t = \phi_1Y_{t-1}+\phi_2Y_{t-2}+\dots+ \phi_{r}Y_{t-r}+u_t,
\end{equation*}
where $u_t$ is independent of past observations. The conditional quantile of the response variable can be directly expressed as a linear combination of $\{Y_{t-j}\}_{j=1,\dots, r}$,
\begin{equation*}
    \begin{split}
    Q_{Y_t}\left(\tau|Y_{t-1},\dots, Y_{t-r}\right) & = Q_{u_t}\left(\tau|Y_{t-1},\dots, Y_{t-r}\right)+\phi_1Y_{t-1}+\phi_2Y_{t-2}+\dots+\phi_{r}Y_{t-r}     \text{ for } \forall \tau \in (0,1)\\
    & = \underbrace{Q_{u_t}(\tau)}_{\theta_0(\tau)}+\underbrace{\phi_1}_{\theta_{1}(\tau)}Y_{t-1}+\underbrace{\phi_2}_{\theta_2(\tau)}Y_{t-2}+\dots+\underbrace{\phi_{r}}_{\theta_r(\tau)}Y_{t-r}.
    \end{split}
\end{equation*}
\subsection{Simulations of Extensions}\label{extensionsimulation}
In this section, we present some simulation results of non-causality testing strategies extended to the autoregressive processes with heteroskedasticity. In this stage, we assume the form of heteroskedasticity is known.\newline
Consider a pair of MAR($1,0$)-ARCH($1$) and MAR($0,1$)-ARCH($1$)  processes defined by
\begin{equation}\label{arch_ar1}
    \begin{cases}
    Y_t = 0.7Y_{t-1}+v_t \\
   Y^*_t = 0.7^{-1}Y^*_{t-1}+v_t = 0.7Y^*_{t+1}-0.7v_{t+1}\\
   v_t=\sigma_t u_t \\
   \sigma_t=0.2 + 0.8\left\vert v_{t-1}\right\vert \text{ where }u_t \sim IID(0, 1) .
    \end{cases}
\end{equation}
As explained in the extension section, we want to detect non-causality by checking whether the coefficients (except the intercept) in the following linear dynamic quantile model are $\tau$-invariant
\begin{equation}\label{qar_arch}
\begin{split}
   Q_{Y_t}\left(\tau|Y_{t-1},v_{t-1}\right)   & = \theta_0(\tau)+\theta_1(\tau)Y_{t-1}+\theta_2(\tau)|v_{t-1}|, \ \ \tau \in \Upsilon \subset (0,1)
    \end{split}
\end{equation}
provided that $v_{t-1}$ is recovered with $100\%$ accuracy.\newline
Another approach is to check whether the linear model (\ref{arch_ar1}) is the correct specification for the conditional quantile of $Y_t$ and $Y^*_t$. The innovation $u_t$ varies from exponential to t student and Laplace distributions. The sample size in this trial is 100, 200, 500, and 1000. The trimmed quantile interval for the constancy test is $[0.05, 0.95]$ and $\Upsilon= [0.01, 0.99]$ for the specification-based test (here in this experiment, we only apply the EV test to see its performance). The empirical size and power of non-causality tests for cases with heteroskedasticity are displayed in Table \ref{tab:ARCH}.
It is clear that both methods have fairly good performance, even in relatively small samples. Regarding the empirical size, both approaches have a slight distortion compared to the nominal level. In the case of the EV test, a different bandwidth can be applied to adjust the empirical size to the expected level. Concerning the empirical power, the specification-based approach dominates the constancy test in most cases. Except in the case of asymmetric distribution, the constancy test outperforms the EV test in small samples ( T=100 and 200 ). It is conceivable that when we replace $v_t$ by its estimate $\hat{v}_t$, the asymptotic effect of the estimation needs to be taken into account when we construct test statistics. This is beyond the scope of this paper.    
\begin{table}[ht]
\centering
 \begin{adjustbox}{width=\textwidth}
 \begin{threeparttable}
\caption  {Empirical size and power of non-causality tests for AR-ARCH models with known heteroskedasticity}
\label{tab:ARCH}
\begin{tabular}{@{} ccccccccccccc @{}}
\toprule
  \multicolumn{1}{c}{Distribution} & test type & \multicolumn{2}{c}{T=100}& &\multicolumn{2}{c}{T=200} & &\multicolumn{2}{c}{T=500} & & \multicolumn{2}{c}{T=1000}  \\ 
\cline{3-4} \cline{6-7} \cline{9-10} \cline{12-13}
       \multicolumn{1}{c}{$u_t$}   &    & size & power &  & size & power  &  & size  & power & & size & power \\    \midrule
 \multirow{2}{*}{Exponential}      & constancy test   & 3.60\% & 39.00\% &  & 4.40\% & 54.60\% &  & 6.00\% & 75.80\% &   & 7.00\% & 90.80\% \\
   & EV test &  5.20\% & 16.80\% &  & 4.40\% & 34.80\% &  & 4.60\% & 82.80\% &  & 5.00\% & 97.80\% \\  \hline        
   \multirow{2}{*}{t student} &  constancy test & 5.20\%  & 24.60\%  &  & 6.60\% & 39.40\% &  & 6.00\% & 63.60\% & & 7.80\% & 78.00\%  \\
  & EV test & 8.20\% & 51.00\% & & 7.20\% & 83.20\% &  & 8.20\% & 99.80\% & & 7.20\% &   100.00\%  \\ \hline
   \multirow{2}{*}{Laplace}      & constancy test   & 4.40\% & 14.40\% &  & 3.80\% & 25.00\%  & & 7.60\% & 38.80\% &  & 4.40\% & 55.00\% \\
   & EV test & 7.00\% & 45.40\% & &  6.40\% & 82.60\% &  & 8.60\% & 99.20\% & & 7.20\% & 10.00\% \\


\bottomrule
\end{tabular}
\begin{tablenotes}
\item The bandwidth for approximating the critical value using subsampling for the EV test is 7. 

\end{tablenotes}
\end{threeparttable}
\end{adjustbox}
\end{table}
\end{document}